\documentclass[aps,prd,reprint,showpacs,superscriptaddress,nobibnotes,nofootinbib]{revtex4-1}
\usepackage{graphicx}
\usepackage{amssymb} 
\usepackage{amsmath}
\newcommand{\vect}[1]{\boldsymbol{#1}}
\begin{document}

\title{Measurement of the Permanent Electric Dipole Moment of the $^{129}$Xe Atom}
\author{F. Allmendinger}
\email{Corresponding author: allmendinger@physi.uni-heidelberg.de}
\affiliation{Physikalisches Institut, Ruprecht-Karls-Universit\"{a}t, 69120 
Heidelberg, Germany}

\author{I. Engin}
\affiliation{Peter Gr\"{u}nberg Institute (PGI-6), Forschungszentrum J\"{u}lich, 
52425 J\"{u}lich, Germany}

\author{W. Heil}
\affiliation{Institut f\"{u}r Physik, Johannes Gutenberg-Universit\"{a}t, 
55099 Mainz, Germany}

\author{S. Karpuk}
\affiliation{Institut f\"{u}r Physik, Johannes Gutenberg-Universit\"{a}t, 
55099 Mainz, Germany}

\author{H.-J. Krause}
\affiliation{Institute of Complex Systems (ICS-8), Forschungszentrum J\"{u}lich, 
52425 J\"{u}lich, Germany}

\author{B. Niederl\"{a}nder}
\affiliation{Institut f\"{u}r Physik, Johannes Gutenberg-Universit\"{a}t, 
55099 Mainz, Germany}

\author{A. Offenh\"{a}usser}
\affiliation{Institute of Complex Systems (ICS-8), Forschungszentrum J\"{u}lich, 
52425 J\"{u}lich, Germany}

\author{M. Repetto}
\affiliation{Institut f\"{u}r Physik, Johannes Gutenberg-Universit\"{a}t, 
55099 Mainz, Germany}

\author{U. Schmidt}
\affiliation{Physikalisches Institut, Ruprecht-Karls-Universit\"{a}t, 69120 
Heidelberg, Germany}

\author{S. Zimmer}
\affiliation{Physikalisches Institut, Ruprecht-Karls-Universit\"{a}t, 69120 
Heidelberg, Germany}

\date{\today}
\begin{abstract}
We report on a new measurement of the CP-violating permanent Electric Dipole Moment (EDM) of the neutral $^{129}$Xe atom. Our experimental approach is based on the detection of the free precession of co-located nuclear spin-polarized $^3$He and $^{129}$Xe samples. The EDM measurement sensitivity benefits strongly from long spin coherence times of several hours achieved in diluted gases and homogeneous weak magnetic fields of about 400~nT.  A finite EDM is indicated by a change in the precession frequency, as an electric field is periodically reversed with respect to the magnetic guiding field. Our result, $\left(-4.7\pm6.4\right)\cdot 10^{-28}$~\textit{e}cm, is consistent with zero and is used to place a new upper limit on the $^{129}$Xe EDM: $|d_\text{Xe}|<1.5 \cdot 10^{-27}$~\textit{e}cm~(95\% C.L.). We also discuss the implications of this result for various CP-violating observables  as they relate to theories of physics beyond the standard model.

\end{abstract}

\maketitle
\section{Introduction and Theoretical Motivation}
Precision measurements of fundamental symmetry violations in atoms can be used as a test of the Standard Model (SM) of particle physics and to search for or to put limits on physics beyond the SM. Permanent Electric Dipole Moments (EDMs) of fundamental or composite particles are excellent candidates to look for new sources of CP symmetry violation, the combined symmetry of charge conjugation C and parity P. CP violation is well known within the SM as a property of the weak interaction and is incorporated (as a complex phase factor) into the CKM matrix describing quark mixing. Since the CP-violating phase enters only where heavy quarks are involved and higher order loops are needed to generate particle EDMs, SM contributions to EDMs are inevitably very small. For example, the SM prediction for the neutron EDM is $d_\text{n}\approx10^{-34}$~\textit{e}cm \cite{Khriplovich}, and for the electron EDM $d_\text{e}\approx10^{-44}$~\textit{e}cm \cite{Pospelov}. Measurements of significantly larger EDMs would be clear indications of additional sources of CP violation (flavor conserving) and Beyond-Standard-Model (BSM) physics. Conversely, to the extent that an EDM is not seen in increasingly sensitive experiments, some BSM scenarios such as the minimal super-symmetric extension of the SM (MSSM), left-right symmetric models and extended Higgs sectors are strongly disfavoured \cite{Chupp}.\\
There are four distinguishable lines of experimental approach in EDM search~\cite{Jungmann}: single free elementary particles and atomic nuclei (\textit{e.g.} neutron (n), electron (e) and muon ($\mu$) ), atoms and ions (\textit{e.g.} mercury (Hg) and xenon (Xe)), molecules and molecular ions (\textit{e.g.} ytterbium fluoride (YbF), thorium oxide (ThO), hafnium fluoride ion (HfF$^+$)), and condensed matter (\textit{e.g.} ferroelectric materials). The observation of an EDM in any system will be a high achievement. However, a single system alone may not solve the questions arising in the connections to the underlying fundamental theory and to cosmology, for example separating weak and strong CP violation. The recent reviews~\cite{Chupp,Chupp2,Engel} cover the experimental approaches in EDM search and the theoretical interpretations of EDM limits. The most precise EDM measurements to date were performed in using neutral particles ($n$)~\cite{Baker}, diamagnetic atoms (Hg)~\cite{Graner1,Graner2}, polar molecules (ThO)~\cite{Andreev} and molecular ions (HfF$^+$)~\cite{Cairncross}.\\
Here, we present the results of an improved EDM search in the diamagnetic $^{129}$Xe atom. The upper limit obtained $|d_\text{Xe}|<1.5 \cdot 10^{-27}~e\text{cm~(95\% C.L.)}$ sets a three times tighter constraint than the recent limit of Sachdeva \textit{et al.}~\cite{Sachdeva} who could slightly improve the 2001 result of Rosenberry \textit{et al.}~\cite{Rosenberry}. EDM experiments can also set new constraints on axion-mediated CP-violating interaction between atomic electrons and the nucleus~\cite{Dzuba3}. From our result, limits for a specific combination of scalar and pseudoscalar coupling constants are derived for the diamagnetic Xe atom. Our method is based on detection of free spin precession of co-located gaseous, nuclear polarized $^3$He and $^{129}$Xe samples. Since this type of a co-magnetometer will preferably be operated at low magnetic fields of about 400 nT, and thus, at low frequencies ($\approx$~10~Hz), using a SQUID as magnetic field detector is appropriate due to its high sensitivity in that spectral range.
\section{Principle of the experiment}
This section gives a short overview of the basic principle of the experiment to measure the EDM of the $^{129}$Xe atom: the neutral $^{129}$Xe atom is a spin-1/2 particle with a corresponding nuclear magnetic moment $\mu_{\text{Xe}}=\frac{1}{2} \hbar \gamma_{\text{Xe}}$, where $\gamma_\text{Xe}$ is the gyromagnetic ratio. If the two-level atom with a non-zero EDM $d_\text{Xe}$ is placed in aligned electric $\vect{E}=\left(0,0,E_{z}\right)$ and magnetic fields $\vect{B_0}=\left(0,0,B_{z}\right)$, the energy splitting is directly proportional to the precession frequency $\omega_\text{Xe}$: 
\begin{eqnarray}\label{eqn:elevel}
\Delta E&=&\hbar \omega_\text{Xe}=|\hbar \gamma_{\text{Xe}} B_{z}+2 d_\text{Xe} E_{z}|~~.
\end{eqnarray}
If the magnetic field is constant, a finite EDM is indicated by the corresponding change in $\omega_\text{Xe}$ as the electric field is reversed. To render the experiment insensitive to fluctuations and drifts of the magnetic guiding field, the principle of co-magnetometry is used: two different spin species are located in the same volume; in our case hyperpolarized $^{129}$Xe and $^3$He gas. The latter has a nuclear spin of $I=1/2$, too, with gyromagnetic ratio $\gamma _\text{He}$. As observable, the weighted frequency difference is used, defined as
\begin{eqnarray}\label{eqn:frequencydiff}
\Delta \omega =\omega_\text{Xe} -\frac{\gamma _\text{Xe} }{\gamma _\text{He} }\omega _\text{He}~~.
\end{eqnarray}
Using Eq.~(\ref{eqn:elevel}), this results in
\begin{eqnarray}\label{eqn:frequencydiff2}
\Delta \omega =\pm \frac{2}{\hbar} d_\text{Xe} |E_{z}|~~.
\end{eqnarray}
The plus sign applies to parallel $\vect{E}$ and $\vect{B}$ fields, the minus sign to the anti-parallel case. Here, the co-located nuclear polarized $^3$He atoms solely serve as a co-magnetometer. EDM contributions in helium are strongly suppressed by Schiff screening ($d\propto Z^{2}$) \cite{Schiff,FlambaumKozlov}. Note that for ideal co-magnetometry, the weighted frequency difference directly projects out the EDM effect one is looking for, without the need to switch the electric field. In addition, the modulation of the $E$-field helps to suppress higher order effects which do not drop out in co-magnetometry. For practical reasons we evaluate Eq.~(\ref{eqn:phasediff}), which is the integrated form of Eq.~(\ref{eqn:frequencydiff}) over time. The weighted phase difference
\begin{eqnarray}\label{eqn:phasediff}
\Delta \Phi =\Phi _\text{Xe} -\frac{\gamma _\text{Xe} }{\gamma _\text{He} }\Phi _\text{He}
\end{eqnarray}
is expected to be constant in the case of pure magnetic interaction. However, non-magnetic spin interactions, like the coupling of the EDM to an electric field, do not drop out. On a closer inspection, the effect of Earth's rotation (\textit{i.e.}~the rotation of the SQUID sensors with respect to the precessing spins) is not compensated by co-magnetometry as well as frequency shifts due to the Ramsey-Bloch-Siegert (RBS) shift. Those effects are discussed in section~\ref{sec:detphaseshifts} and have to be accounted for in the data evaluation.\\
In this experiment, the precession of the transverse sample magnetization of $^3$He and $^{129}$Xe is monitored. A finite EDM is indicated by a corresponding change in $\Delta \omega$ as the electric field is reversed. The statistical sensitivity to determine frequency changes is given by the Cramer-Rao Lower Bound (CRLB) \cite{Gemmel,Kay}. The statistical uncertainty of the EDM measurement $\sigma_d$ is proportional to
\begin{eqnarray}\label{eqn:sensitivity}
\sigma_d &\propto& \frac{\rho\sqrt{C\left(T/T_2^*\right)}}{E_z A_0 T^{3/2} }~~,
\end{eqnarray}
where $T$ is the measurement time of coherent spin precession, $C\left(T/T_2^*\right)$ describes the effect of exponentially damped sinusoidal signal with amplitude $A_0$, and $\rho$ is the noise level at the relevant frequencies.
According to Eq.~(\ref{eqn:sensitivity}), the following conditions should be met in order to achieve a high resolution EDM measurement: \\
i) Long transverse spin coherence times, the characteristic time constant given by $T_2^*$. Due to the $T^{3/2}$ behavior, the experiment strongly benefits from long $T_2^*$ of several hours which are achievable in diluted gases with magnetic field gradients in the 10~pT/cm range~\cite{Gemmel}.\\
ii) A high electric field $E_z$ across the spin sample.\\
iii) A high signal-to-noise ratio (SNR$=A_0/\rho$), \textit{i.e.} a high signal $A_0$ and a low noise level $\rho$ at the relevant frequencies.\\
The key to an improved EDM sensitivity is the reduction of magnetic field gradients, as they directly and indirectly influence the relevant system parameters which determine the EDM sensitivity (Eq.~(\ref{eqn:sensitivity})): according to~\cite{Cates}, the transverse relaxation time is given by $1/T_2^*= 1/T_1+1/T_\text{2,grad}$ with $T_\text{2,grad} \propto D/|\nabla \vect{B}|^2$. Assuming the longitudinal relaxation time $T_1$ to be sufficiently long (see Section \ref{sec:EDMcelldesign}), we have a direct quadratic dependence of $T_2^*$ on the absolute field gradients. The dependence on the diffusion coefficient $D$ suggests to measure at  low gas pressures ($D\propto1/p$ ). As a result, the signal amplitude $A_0$ decreases to the same extent as well as the field strength $E_\text{B}$ at which dielectric breakdown occurs (Paschen curve~\cite{Paschen}) which in turn sets limits for the strength of the applied electric field $E_z < E_\text{B}$. Therefore, the approach in our case is to minimize magnetic field gradients which then provides a higher flexibility in the parameter settings to improve the  statistical uncertainty of the EDM measurement.

\section{Experimental setup and technique}
\begin{figure}
\includegraphics[width=\columnwidth]{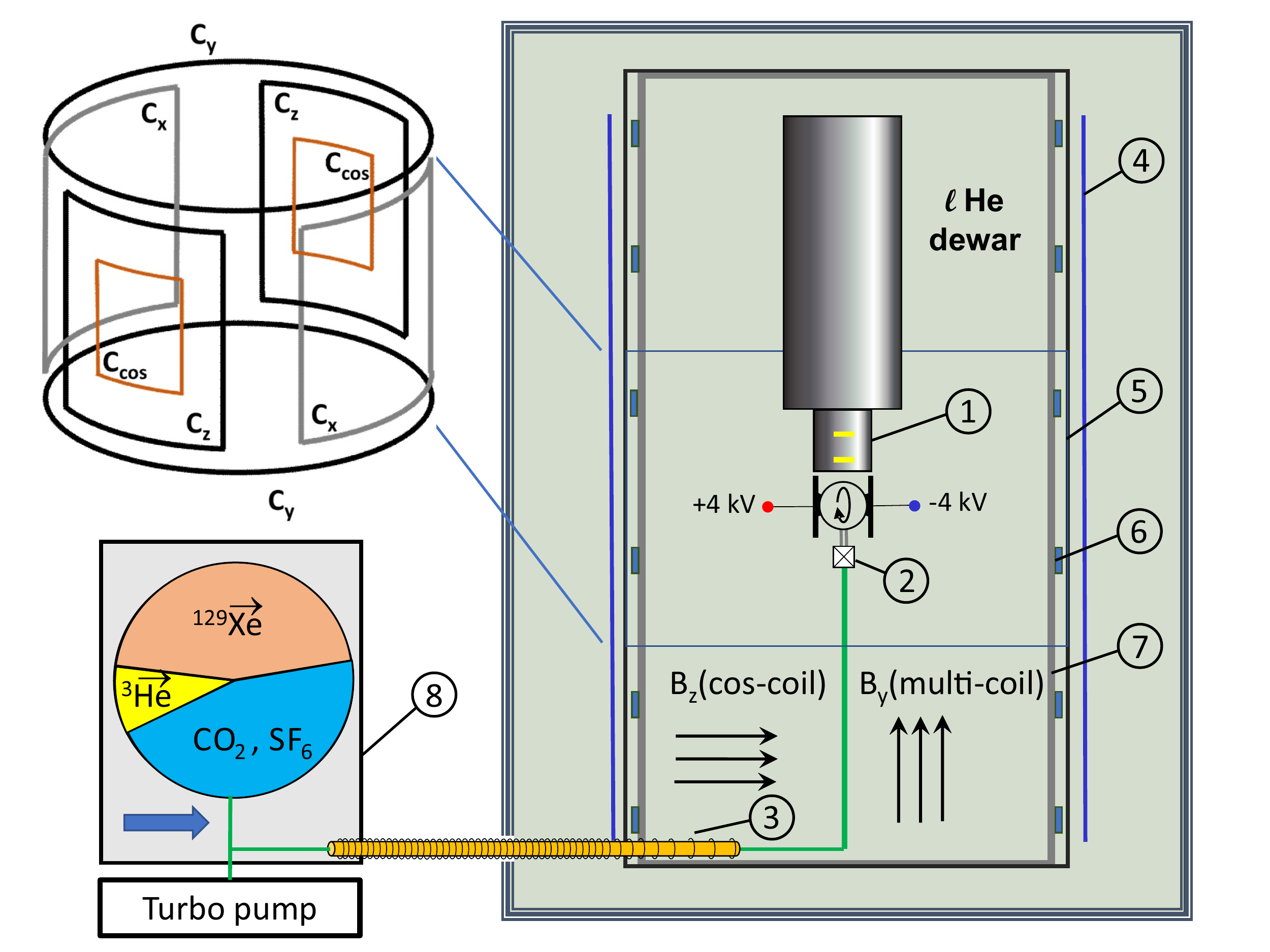}
\caption{Schematic view of the EDM experiment setup. The central part of the experiment, \textit{i.e.} the SQUID-gradiometer (1) and the EDM cell (2), is placed inside a two-layer magnetically shielded room (MSR) with an additional mu-metal cylinder (4) to reduce magnetic field gradients. A coil assembly consisting of a cosine-coil (5) and an axial multi-coil system (6) generates a homogeneous magnetic guiding field in transverse and longitudinal direction, respectively. Four additional shimming coils (shown in the top-left corner) are used to compensate gradients. A fibre-reinforced plastic tube (7) acts as a rigid mounting structure for all components inside the MSR. The gas mixture with the hyperpolarized noble gases $^3$He and $^{129}$Xe and additional buffer gases is provided outside the MSR (8). By means of a gas-transfer system the gas mixture is expanded into the pre-evacuated EDM cell. Solenoidal coils along the transfer line with decreasing winding number density (3) ensure an adiabatic spin transfer from the outer holding field of a few $100~\mu$T to the low field region inside the MSR. Demagnetization coils around the MSR and the mu-metal cylinder (not displayed) are used to obtain reproducible low field gradients. A more detailed view of the EDM cell is given in Fig.~\ref{fig:cell}.
\label{fig:setup}
}
\end{figure}

The individual components and procedures of the experiment are presented in the following section. Figure~\ref{fig:setup} gives a schematic overview of the setup while a more detailed view on the EDM cell assembly is shown in Fig.~\ref{fig:cell}.
\subsection{Magnetic shielding and coil system}
The experiment is placed inside a magnetically shielded room (MSR) at the Institute of Complex Systems, Research Center J\"ulich, Germany. The MSR consists of two layers of mu-metal with a wall thickness of 1.27~mm each, and a high frequency shield of 10~mm  aluminum. The inner dimensions of the walk-in MSR are $3.00~\text{m} \times 2.50~\text{m} \times 2.35~\text{m}$. An additional mu-metal cylinder (diameter 0.85~m, height 1.9~m, wall thickness 1.5~mm) is placed centrally inside the MSR to reduce the existing magnetic field gradients from 300~pT/cm to 50~pT/cm in a first step. The gain in spin-coherence time by reducing the field gradients in the vicinity of the EDM cell overcompensates the noise-level increase from $\approx$1 to 10~fT/$\sqrt{\text{Hz}}$  (see Fig.~\ref{fig:spectrum}) due to the elevated Johnson noise generated by this high-permeability magnetic shield \cite{Lee}. Both the MSR and the mu-metal cylinder are equipped with demagnetization coils. \\
The central parts of the EDM experiment (\textit{i.e.} the EDM cell containing the hyperpolarized gases and the SQUID-magnetometer system) are placed inside the mu-metal cylinder, as well as the coil system that generates the homogeneous magnetic guiding field. A fibre-reinforced plastic tube acts as a rigid mounting structure for all devices, effectively suppressing low-frequency vibrations of the individual components relative to each other. The tube itself is fixed to the frame structure of the MSR with built-in vibration damping materials. This measure reduces interfering vibration modes in the low frequency range (1-30 Hz) seen by the SQUID system as it moves through existing magnetic gradient fields. A cosine-coil with a diameter of 0.8~m and a length of 2.1~m produces a homogeneous magnetic field inside the cylinder perpendicular to the cylinder axis. Removable printed circuit boards form the top and bottom lids of the cosine coil, allowing access to the inner parts of the experiment~\cite{Zimmer}. In addition, a uniform magnetic field along the cylinder axis generated by a multi-coil system serves for spin manipulation. In order to reach the required long transverse relaxation times of several hours, it is necessary to further minimize the magnetic field gradients: four additional shimming coils along the cylinder axis (Anti-Helmholtz coils) and in transverse direction (saddle coils) are used to actively compensate the $\approx$50 pT/cm gradient fields inside the innermost shield at the position of the EDM cell~\cite{Grasdijk}~(see Fig.~\ref{fig:setup}).\\
Very stable and adjustable low-noise current sources drive the coil system. The output current is programmable from -50 to 50~mA with a resolution of $\Delta I=100$~nA and a maximum frequency of 1~kHz. In order to avoid conducting noise from the environment into the MSR, the current sources are controlled from outside via an optical link and are powered by batteries; a scheme that is maintained for all electronic devices in the setup.
\subsection{SQUID gradiometers and data acquisition} 
Superconducting Quantum Interference Devices (SQUIDs) are used to measure the precessing $^3$He and $^{129}$Xe magnetization. The low-temperature DC-SQUID gradiometer system made by Magnicon~\cite{Magnicon} reaches an intrinsic noise level of 0.7~fT/$\sqrt{\text{Hz}}$ above the $1/f$-noise limit of 1~Hz. Two loops with a diameter of 30 mm separated axially by a distance 70 mm and connected in series opposition form a first-order axial gradiometer. The loops are transformer-coupled to the SQUID. The SQUID itself is shielded from any external magnetic field by a niobium capsule. Thus, readings from far away sources and ambient magnetic noise will be suppressed by a factor called the common mode rejection ratio. However, signal sources next to the lower gradiometer loop with a typical dipole-field distribution are attenuated very little.  The SQUID system is placed inside a liquid helium cryostat manufactured by Cryoton \cite{Cryoton}. The low magnetic noise fiberglass model was tested to be free of magnetizable material (\textit{e.g.} small ferromagnetic particles). The distance between the inner volume at liquid helium temperature and the outside at room temperature is 14~mm. The inner volume (about 16 liters) is filled with liquid helium which keeps the lower part of the cryostat cold for about one week without refilling. The room-temperature part of the SQUID readout electronics is placed on top of the cryostat. As this experiment is based on precision measurements of signal phases, special care has to be taken to avoid non-linear phase shifts that depend on frequency or temperature. Such phase shifts can easily occur when using simple RC low-pass filters for anti-aliasing, for instance. Therefore, the analog output signals are digitized by delta-sigma ADCs \cite{ADS1299} which effectively sample the input at a high frequency (here, $1.024$~MHz). This allows for a high frequency low-pass anti-aliasing filter with negligible phase shifts at the relevant helium and xenon Larmor-frequencies (roughly 5 and 13~Hz at the chosen magnetic holding field of about 400~nT). The advantage of delta-sigma ADCs is that most of the conversion process is implemented in the digital domain and very few analog components are needed. This results in a high performance with respect to noise and phase shifts. The ADC sampling rate is adjustable. In our case, it was set to 250~Hz.
\subsection{EDM cell design}
\label{sec:EDMcelldesign}
\begin{figure}
\includegraphics[width=\columnwidth]{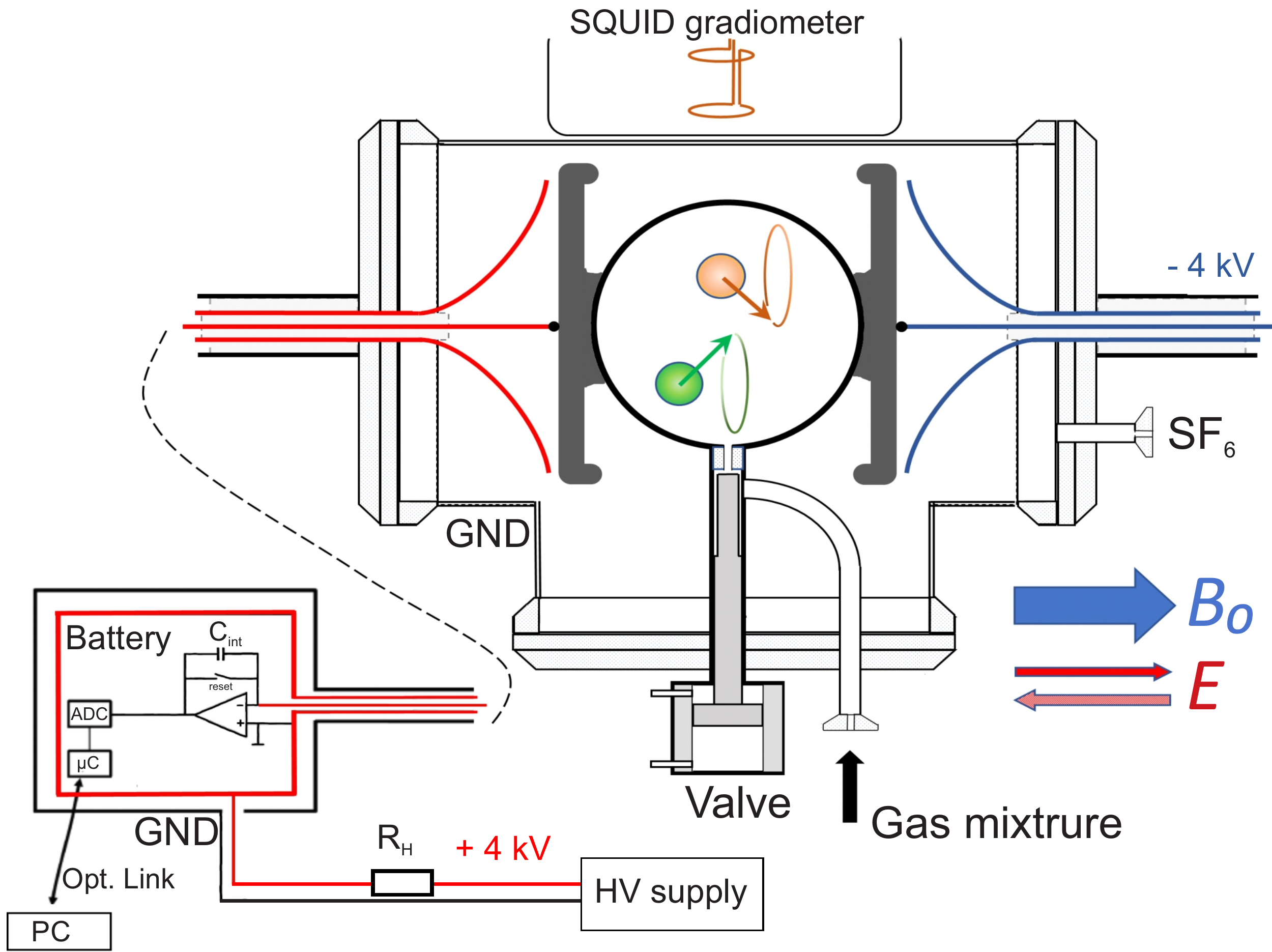}
\caption{Schematic view of the spherical EDM cell in the homogeneous electric field of two plate-capacitor electrodes. The setup itself is inside a T-shaped enclosure which is flooded with SF$_6$ for dielectric insulation. Double-shielded cables serve as +/- HV supply lines (details see text). Leakage currents are measured by insulated pA-meters put on the respective +/- potentials. The HV supply itself is positioned outside the MSR. Funnel-shaped electrodes at the same potential as the inner +/- HV supply lines further prevent leakage currents to the grounded conductive walls (carbon coated) of the glass T-piece. By means of a SQUID gradiometer (on top) the transverse magnetization of the precessing spin sample is monitored. The hyperpolarized gas mixture enters the cell volume through a pneumatically driven valve.
\label{fig:cell}
}
\end{figure}
A prerequisite to reach long spin coherence times are measurement cells which show low wall relaxation rates ($1/T_\text{1,wall}$)  for both hyperpolarized gases. The  EDM cell is a spherical cell with an outer diameter of 100~mm, completely made of GE-180 glass. As demonstrated in \cite{Repetto}, wall relaxation times of almost 20~h can be achieved for $^{129}$Xe, while more than 100~h have been reported for $^3$He \textit{e.g.} in \cite{Schmiedeskamp,Rich}. Carbon-coated (conductive) glass electrodes arranged in form of a plate capacitor directly touch the outer wall of the spherical EDM cell. They are aligned in such a way that the electric field is oriented parallel to the magnetic guiding field of the cosine coil ($z$-direction). Additional shielding electrodes (carbon-coated glass) at the same potential to a certain extent prevent leakage currents to the environment, \textit{i.e.}, to an encasing T-shaped glass tubing (carbon-coated) held at ground potential. The housing is repeatedly flooded with SF$_6$ to prevent sparking. The use of external electrodes to define a homogeneous electric field across the EDM cell has two reasons: a) to reach long spin-coherence times that are not limited by a faster wall relaxation caused by the electrode material (\textit{e.g.} silicon), and b) to circumvent demagnetization effects which lead to enhanced Ramsey-Bloch-Siegert phase shifts in case of  imperfect spherical symmetry of the spin sample by using internal electrodes or by the choice of cell geometries other than spherical ones, \textit{e.g.} cylindrical cells (see Section \ref{sec:detphaseshifts}). 
A pneumatically driven valve made of PEEK allows a remote controlled opening and closing of the glass cell via its short-stemmed inlet/outlet port. This way, deviations from spherical symmetry are kept as small as possible when the cell is filled with the hyperpolarized gas mixture.
\subsection{Electric field generation and leakage current monitors}
A high precision dual channel high voltage module (NHQ by Iseg company~\cite{Iseg}) is used for the electric field generation. One channel is permanently set to positive output (adjustable from 0 to +6~kV) and the other one to negative output (0 to -6 kV). The output voltages and currents can be monitored remotely with a resolution of 100~mV and 100~pA. Four high voltage relays are used to select the negative or positive voltage supply individually for each EDM-cell electrode. The ripple of the NHQ-output voltage (less than 5~mV peak to peak) is further reduced by RC low-pass filters. High-impedance resistors ($R_\text{H}=100~\text{M}\Omega$) at the output prevent large currents, \textit{e.g.} in the case of sparking. The high voltage supply and the relays are placed outside the MSR in order to avoid magnetic effects correlated with the switching of the relays. The high voltage is fed into the MSR by high resistance conductors (several M$\Omega$) to minimize noise inside the EDM setup.\\
Currents associated with the high voltage setting give rise to systematic errors (see Section \ref{sec:syseffects}). Therefore, currents that flow in the proximity of the sample cell, especially between the two electrodes, have to be monitored precisely on the pA level. Since cable-leakage currents cannot easily be separated from currents that flow across the EDM cell, the principle of a doubled shielded cable was applied to measure leakage currents in the proximity of the sample cell: the inner wire (carbon mesh) which contacts the electrode and keeps it at the applied  potential, is surrounded by an insulating silicon tube which is shielded by a tubular carbon mesh kept at the same potential. This unit is fitted into a second silicon tube enclosed again by a carbon-mesh shield at ground potential. The two Picoampere-meters (pA-meters) are connected to the respective electrodes with double-shielded cables according to the wiring diagram shown in Fig.~\ref{fig:cell}. By this measure, the pA-meters only monitor leakage currents between the two plate-capacitor electrodes and from the electrodes to the grounded casing.\\
The pA-meters are based on the integrator chip IVC102 (Burr-Brown/Texas Instruments) with a low bias-current precision operational amplifier and various integration capacitors on chip. As the current through the innermost wires has to be measured, the pA-meters have to be put at the high potential. To do so, the pA-meter circuit boards and batteries are placed in an aluminum box. This conductive box is surrounded by an insulating plastic housing to keep it at high potential with respect to the environment which is at ground potential. The pA-meters are read out via an optical interface. The inner shielding of the double shielded cable is directly connected to the aluminum housing of the pA-meter, whereas the innermost wire connects the input of the pA-meter with the electrode of the cell.

\subsection{Hyperpolarization of  $^3$He and $^{129}$Xe, and gas preparation}
$^3$He is hyperpolarized by Metastability Exchange Optical Pumping (MEOP) at the Institute of Physics, University of Mainz using the existing $^3$He polarizing facility \cite{Karpuk} where nuclear polarization degrees above 70\% can be reached \cite{Wolf}. The hyperpolarized  $^3$He gas at a pressure of 1.5 bar is then transferred to the experiment location in low-relaxation glass vessels inside magnetized transport boxes for housing polarized spins in homogeneous fields \cite{Hiebel,Thien}.
 The Xe gas (enriched to 91 \% $^{129}$Xe) is hyperpolarized on site by means of Spin Exchange Optical Pumping (SEOP)\cite{Appelt}. Gas mixtures including buffer gases like N$_2$, CO$_2$ or SF$_6$ needed to suppress the Xe nuclear spin relaxation due to the formation of van der Waals molecules \cite{Repetto} are prepared next to the MSR in a dedicated filling station \cite{Zimmer}. From there, the gas mixture is transferred into the MSR while preserving the polarization (see Fig.~\ref{fig:setup}).
\subsection{Technique: Demagnetization and gradient optimization}
In order to minimize magnetic field gradients, the mu-metal of the MSR, and afterwards the inner mu-metal cylinder, have to be demagnetized after closing the setup. This is always the case after the door of the MSR has to be opened to refill the cryostat, for example. Demagnetization procedures of MSRs which lead to reproducible low residual field gradients are described elsewhere \cite{Thiel,Altarev}. In practice, this is obtained by the application of a slowly alternating (\textit{e.g.}, a sinusoidal) magnetic field in the demagnetization coils whose amplitude decreases according to the chosen envelope function. We used a sequence of exponentially decaying sinusoidal currents with 3~Hz, then at 1~Hz through the demagnetization coils. Each routine lasted 300~s, corresponding to ten characteristic time constants. After that, we obtained satisfactory results with gradients in the order of 50~pT/cm. The white system noise seen by the SQUID gradiometers could be reduced by 40\% reaching $\approx10~\text{fT}/\sqrt{\text{Hz}}$  by performing an additional demagnetization routine at the inner shield directly afterwards with AC currents of 1kHz (200 s duration) and repeating the 3~Hz and 1~Hz demagnetization cycle~\cite{Zimmer}.  The following in-situ method is used to further reduce the magnetic field gradients in the vicinity of the EDM cell: The EDM cell is filled with approximately 30~mbar of hyperpolarized $^3$He. After a non-adiabatic spin flip, the Larmor precession signal is monitored. The transverse relaxation time $T_{2, \text{He}}^*$ is maximized by systematically varying the coil currents of the four shimming coils according to a downhill simplex algorithm \cite{NelderMead}. For each setting of coil currents, $T_{2, \text{He}}^*$ is measured for at least ten minutes. The fully automated optimization procedure takes several hours, improving $T_{2, \text{He}}^*$ from 7500~s to 40000~s. This measure finally led to a reduction of gradients from 50~pT/cm to below 10~pT/cm. In \cite{Allmendinger}, we described the precise measurements of magnetic field gradients extracted from transverse relaxation rates of precessing spin samples. This method has the advantage that an EDM-measurement run can directly follow the gradient optimization procedure without any modifications of the setup (like opening the magnetic shield, for instance).
\subsection{Technique: Procedure of an EDM run}
The individual steps to perform a single EDM-measurement run are: a gas mixture of hyperpolarized $^3$He and  $^{129}$Xe including buffer gases is prepared and filled into a storage/transport cell which is attached to the junction piece of the gas-transfer line to the inside of the MSR. For the gas transfer, the solenoids around the transfer line are switched on, as well as the cosine-coil ($z$-axis). Typical partial pressures in the EDM cell after a remote-controlled triggered expansion of the gas mixture are: $p_\text{He}\approx30$~mbar and $p_\text{Xe}\approx100$~mbar. Then, the magnetic guiding field of the EDM setup and with it the sample spins are slowly rotated (adiabatically) into the vertical direction ($y$-direction). A non-adiabatic field switching back to the default z-direction starts the spin precession in the ($x$,$y$)-plane. Thereafter (after at least 300~s), the high voltage is ramped up with 25~V/s to its maximum value of +/-~4~kV (the initial polarity of the electric field across the sample was varied from run to run). After $T_a/4$, the electric field is inverted by ramping the HV back to zero, switching the relays that define the field-polarity, and ramping the HV up again. Afterwards, the electric field is regularly inverted after the time $T_a/2$ (see Fig.~\ref{fig:EDMfunctions}). This particular pattern of electric field switching was chosen in order to minimize parameter correlation. The SQUID signal and the pA-meter data are recorded for off-line evaluation. After $T\approx3\cdot T_{2,\text{Xe}}^*$, the Xe-signal amplitude has decreased substantially and the particular EDM run is stopped. The characteristic transverse relaxation time of $^3$He, $T_{2,\text{He}}^*$, was typically a factor of 6 longer than $T_{2,\text{Xe}}^*$. Depending on the achieved $T_{2,\text{Xe}}^*$ times, the full period $T_a$ of $E$-field switching varied between 12000~s and 18000~s.
\begin{figure}
\begin{flushright}
\includegraphics[width=\columnwidth]{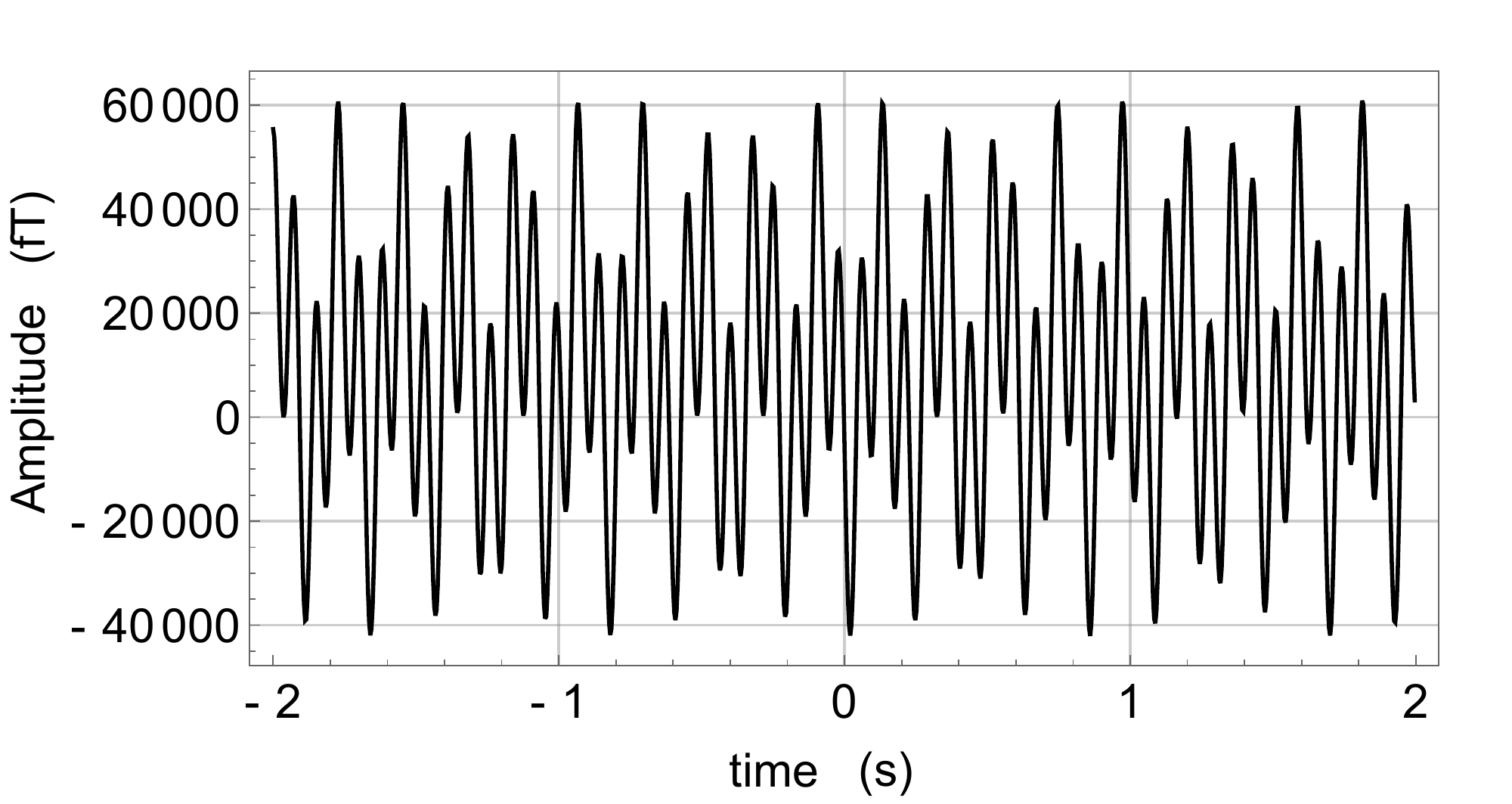}
\includegraphics[width=\columnwidth]{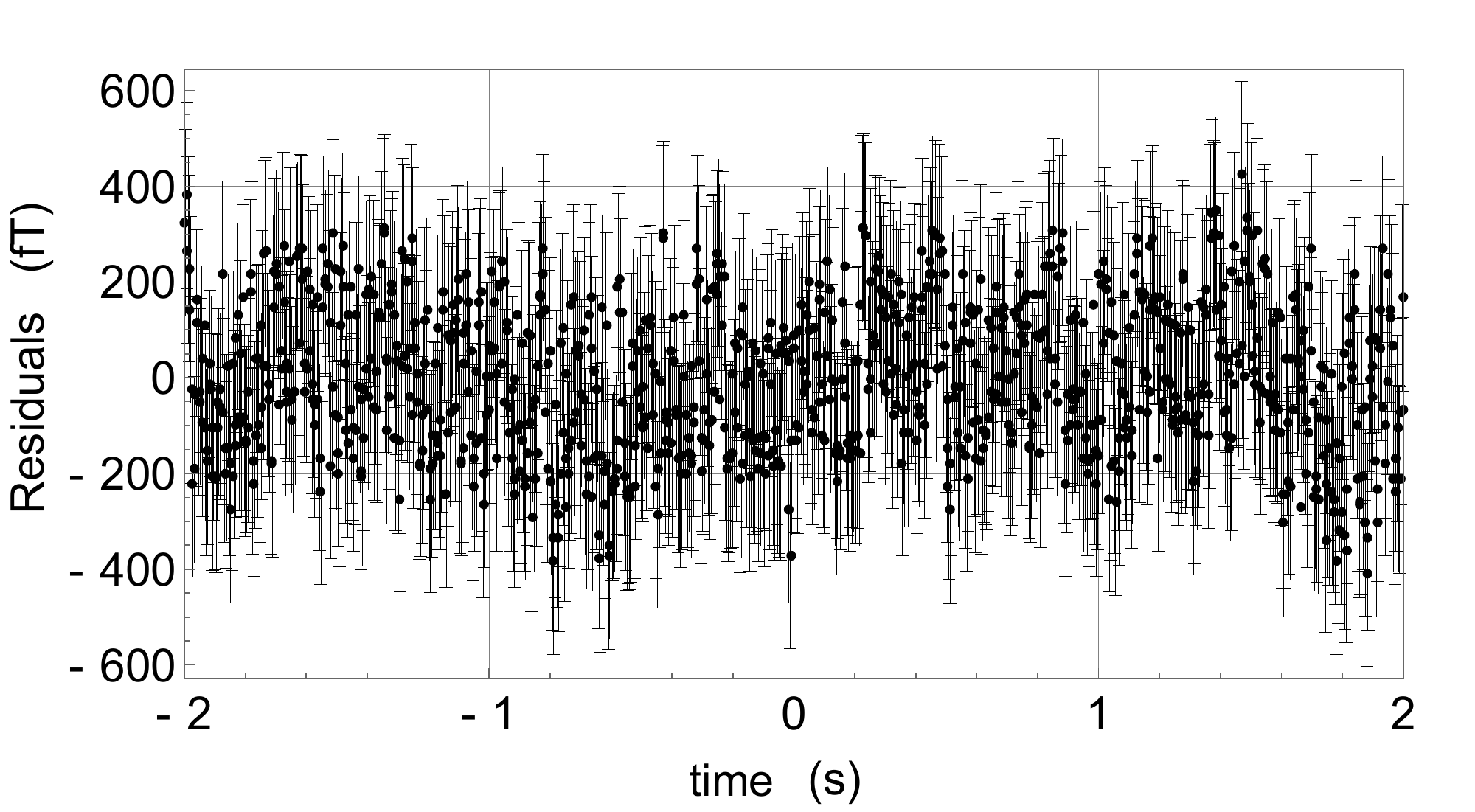}
\end{flushright}
\caption{Top: The SQUID gradiometer raw signal with the prominent beating of the $^3$He and $^{129}$Xe precession signal at the Larmor frequencies $\approx13$~Hz and $\approx5$~Hz (this corresponds to an applied $B_0$~field of about 400~nT). Bottom: The raw signal after subtraction of the fitted model in Eq.~(\ref{eqn:fitrawsignal}) (residuals).
\label{fig:rawsignal}}
\end{figure}
\section{Data evaluation}
In this section, the general data-evaluation procedure with the different steps from raw data to the weighted phase difference and other important intermediate data (signal amplitudes, relaxation time constants, etc.) is discussed. Subsequently, the fit to the weighted phase-difference data in order to extract an EDM value is presented. 
\subsection{Fit to sub-cut data}
\begin{figure}
\includegraphics[width=\columnwidth]{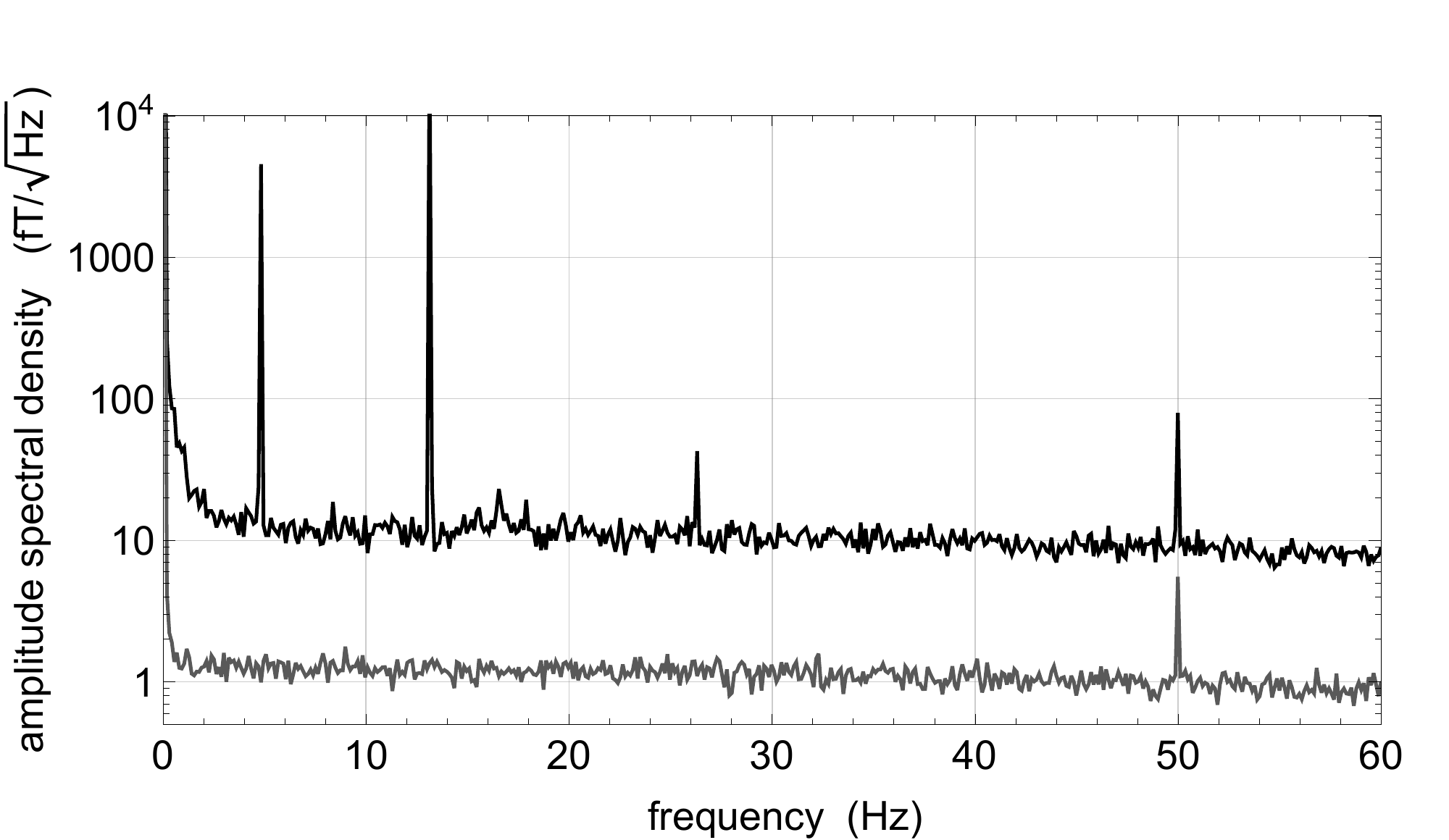}
\caption{The spectrum (amplitude spectral density) of the gradiometer data (black). The prominent sharp peaks at around $5$~Hz and $13$~Hz correspond to the precession frequencies of $^{129}$Xe and $^3$He at $B_0\approx 400$~nT. The narrow peaks at $50$~Hz and $16.6$~Hz are caused by power line interference whereas the peak at 26~Hz is a harmonic component of the strong $^3$He line. For frequencies above 2~Hz we find a white noise level of the gradiometer signal slightly above $\rho\approx$10~fT/$\sqrt{\text{Hz}}$. Without inner mu-metal shield (reduced Johnson noise) the system noise drops by about a factor of 10, reaching $\approx$1~fT/$\sqrt{\text{Hz}}$ which is close to the intrinsic SQUID noise of 0.7~fT/$\sqrt{\text{Hz}}$.
\label{fig:spectrum}}
\end{figure}
To extract the $^3$He and $^{129}$Xe amplitudes, frequencies and phases, the method of piecewise fitting to the gradiometer signal data was applied: the data was split into sets (sub-cuts) with the length of $\Delta t=4$~s. This corresponds to 1000 data points at a sampling rate of $250$~Hz. A typical sub-cut is shown in Fig.~\ref{fig:rawsignal} (top). The assigned uncertainty to each data point is 160~fT. This  value is the typical noise signal derived from the mean system noise $\rho\approx10~\text{fT/}\sqrt{\text{Hz}}$ within the recorded effective bandwidth of 125~Hz (Nyquist frequency). Subsequently, the function
\begin{eqnarray}
\label{eqn:fitrawsignal}
\nonumber f_\text{raw}(t) &= &a_{\text{He}} \cdot \cos \left(\omega _{\text{He}} t\right)+ b_{\text{He}}\cdot\sin \left(\omega _{\text{He}} t\right)\\
\nonumber&&+ a_{\text{Xe}}\cdot \cos \left(\omega _{\text{Xe}} t\right)+ b_{\text{Xe}}\cdot\sin \left(\omega _{\text{Xe}} t\right) \\
&& + c +d\cdot t
\end{eqnarray}
was fitted to the data of each sub-cut. The $\sin$ and $\cos$ terms describe the $^3$He and $^{129}$Xe precession signals at the corresponding Larmor frequencies $\omega _\text{He}$ and $\omega _\text{Xe}$, while the constant and linear terms account for the SQUID offset and a small drift of this offset in time. To minimize the correlation between the constant, linear, $\sin$ and $\cos$ terms, $t=0$ was chosen to be in the middle of the sub-cut, so that the data points are positioned symmetrically around zero from $t=-2$~s to $t=2$~s. The sum of sin and cos terms is chosen to have linear fitting parameters (except $\omega _{\text{He}}$ and $\omega _{\text{Xe}}$) with orthogonal functions. Within the relatively short time interval of the sub-cuts  the term $c +d\cdot t$ represents the adequate parametrization of the SQUID gradiometer offset showing a small linear drift due to the elevated $1/f$-noise at low frequencies (below 1~Hz). On the other hand, the chosen time intervals are long enough to have a sufficient number of data points (1000) for the $\chi^2$ minimization.  Finally, for each sub-cut, one gets a set of estimations for the eight fit parameters $a_\text{He(Xe)}$, $b_\text{He(Xe)}$, $\omega_\text{He(Xe)}$, $c$, and $d$ and their uncorrelated and correlated uncertainties, and, additionally, $\chi^2$ as a measure of the goodness of the fit. The residuals (the measured data after subtraction of the fitted function in Eq. (\ref{eqn:fitrawsignal})) are shown in Fig. \ref{fig:rawsignal} (bottom).\\
\begin{figure}
\includegraphics[width=\columnwidth]{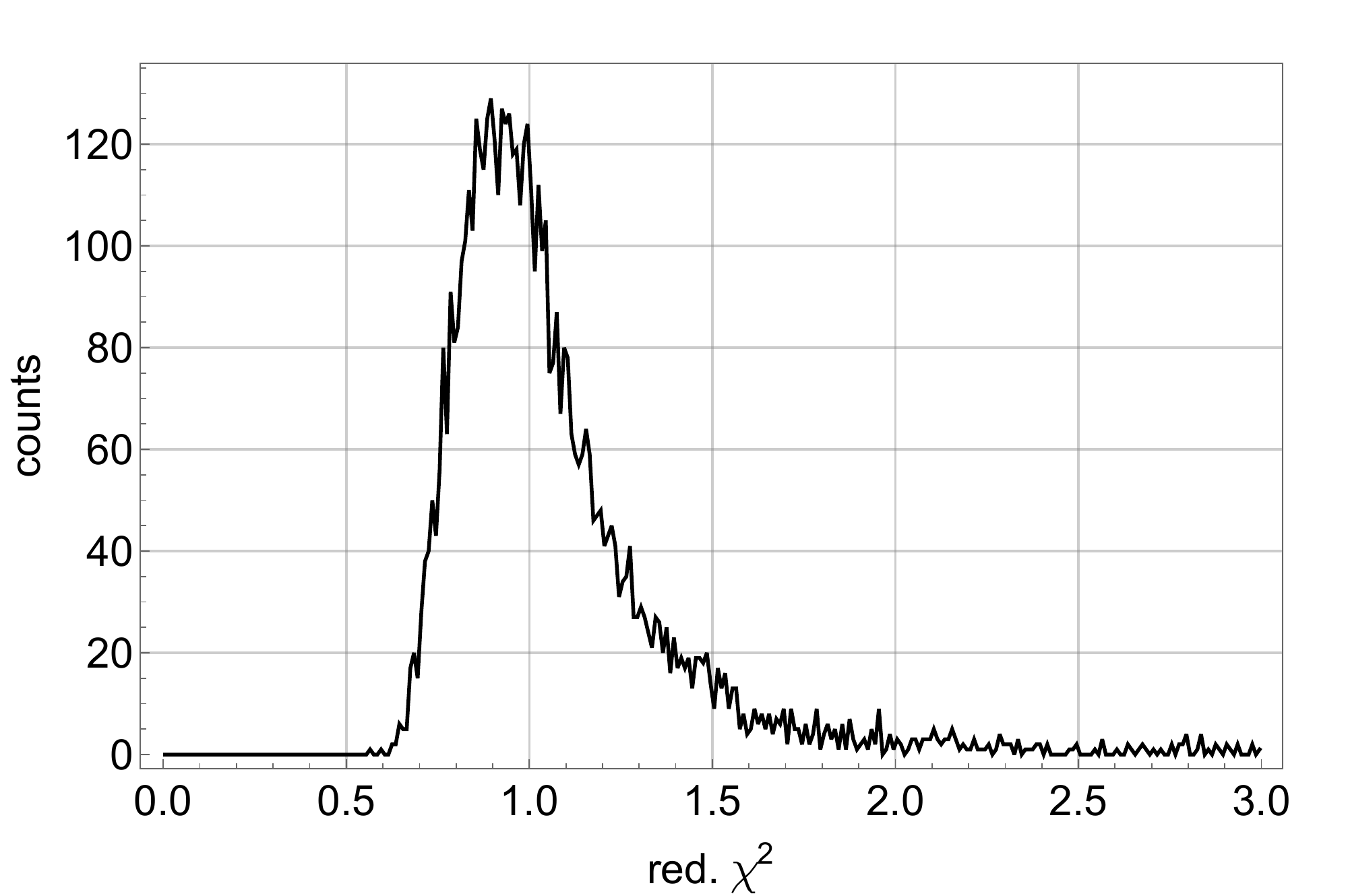}
\caption{The distribution of the sub-cut reduced $\chi^2$ values ($=\chi^2/\nu$ with degrees of freedom $\nu=992$). Here, the data of run number 4 is shown which has 5743 sub-cuts.
\label{fig:chi2}}
\end{figure}
For a measurement run lasting several hours, the number of sub-cuts is in the order of $N\approx 10^4$. In Fig.~\ref{fig:chi2}, the observed reduced-$\chi^2$-distribution of the sub-cut-data fits is displayed. The observed width of the distribution is about a factor of 4  larger than the expected one. This is due to the fact that non-Gaussian noise, \textit{i.e.} higher order components of the $1/f$-noise at low frequencies, as well as slow drifts of the white noise level, have not been included in the fit model. The uncertainties of the extracted fit parameters were scaled with $\sqrt{\chi^2/\nu}$ whenever probability $p(\chi^2 )\le0.05$ was met according to the PDG guidelines~\cite{Beringer}\footnote{$p(\chi^2 ):=\int_{\chi^2}^\infty \text{pdf}(z;\nu) \text{d} z $ where pdf is the probability density function of the $\chi^2$-distribution.}.\\
\begin{figure}
\includegraphics[width=\columnwidth]{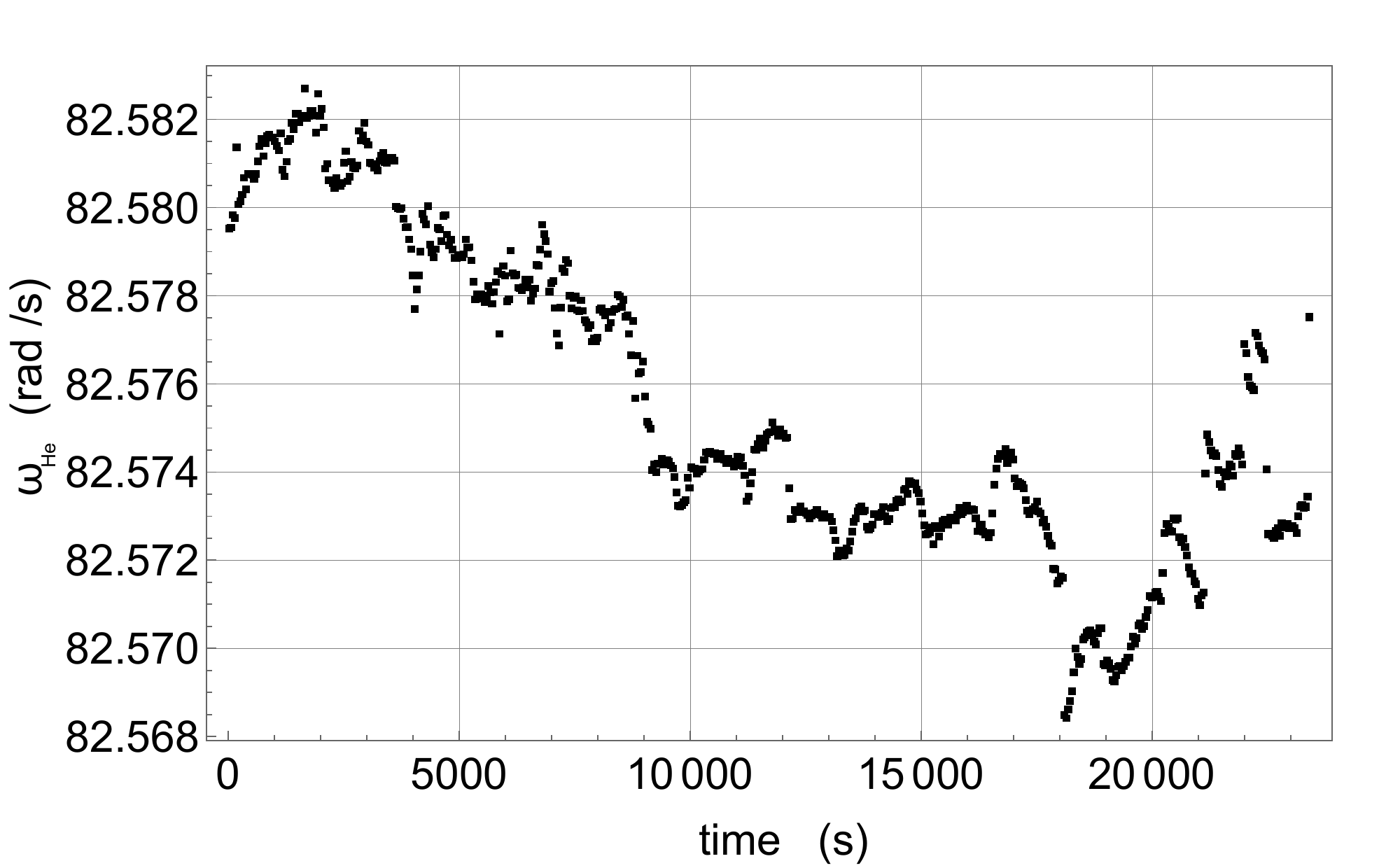}
\caption{The measured Larmor frequency of helium $\omega_{\text{He}}$ as a function of time for measurement run number 6 lasting about seven hours. The time bin per data point is 40~s and the errors are smaller than the symbol size.
\label{fig:omega_He}}
\end{figure}
The fitted Larmor frequencies can be used as a measure for the stability of $B_0$. In Fig. \ref{fig:omega_He}, the measured Larmor frequency $\omega_{\text{He}}$ is plotted as a function of time for measurement run number 6 lasting about seven hours. The relative drift of the magnetic guiding field is in the order of $10^{-5}$ per hour, corresponding to an absolute drift of $10$~pT per hour.
\subsection{Determination of Amplitudes and Phases}
The amplitudes $A_\text{He(Xe)}$ of the $^3$He and $^{129}$Xe signals are calculated from the fit parameters $a_\text{He(Xe)}$ and $b_\text{He(Xe)}$ according to
\begin{eqnarray}
A_\text{He(Xe)}&=&\sqrt{a_\text{He(Xe)}^2+b_\text{He(Xe)}^2}~.
\end{eqnarray}
The transverse relaxation times $T_\text{2, He(Xe)}^*$ are extracted by exponential fits to the amplitude data. As mentioned earlier, $T_2^*$ strongly depends on the gradients of the magnetic field. These gradients are sufficiently constant over the period of a single measurement run, so that the transverse relaxation times can be considered as constant, too. For the further evaluation, the phases of the $^3$He and $^{129}$Xe signals are of main interest as they can be determined very precisely. The phases $\varphi_\text{He}$ and $\varphi_\text{Xe}$ in the range $]-\pi,+\pi]$ of each sub-cut interval being referred to $t=0$ (middle) are determined by
\begin{eqnarray}
\varphi_\text{He(Xe)}&=&\text{arctan2}(b_\text{He(Xe)},~a_\text{He(Xe)} ) ~~.
\end{eqnarray}
The accumulated helium and xenon phases $\Phi_\text{He}$ and $\Phi_\text{Xe}$ are then determined by adding appropriate multiples of $2\pi$. The accumulated phases  increase almost linearly in time (as the Larmor frequencies are almost constant), and after seven hours of measurement, reach about $\Phi_\text{He}(t=7~\text{h})\approx 2.1\cdot10^6\text{~rad}$ and $\Phi_\text{Xe}(t=7~\text{h})\approx 0.8\cdot10^6\text{~rad}$, respectively. The uncertainties (within the 4~s time bin of a sub-cut) of the accumulated phases for $^3$He are on the $0.1$~mrad level, while the corresponding uncertainties for $^{129}$Xe increase from $0.1$~mrad at the beginning to $10$~mrad at the end of the measurement run due to the faster decay of the $^{129}$Xe signal amplitude.\\
Subsequently, the weighted phase difference can be computed according to Eq.~(\ref{eqn:phasediff}) which eliminates the Zeeman-term and only phase shifts due to non-magnetic spin interactions like the coupling of an finite EDM to an electric field remain. On a closer inspection, there are several effects that are not compensated by co-magnetometry: the effect of Earth's rotation (\textit{i.e.} the rotation of the SQUID detectors with respect to the precessing spins), chemical shift, as well as phase shifts due to the Ramsey-Bloch-Siegert (RBS) shift \cite{Bloch, Ramsey}. These different effects lead to deterministic phase shifts. Their origins and time dependencies are described in the following subsection.
\begin{figure}
\begin{flushright}
\includegraphics[width=\columnwidth]{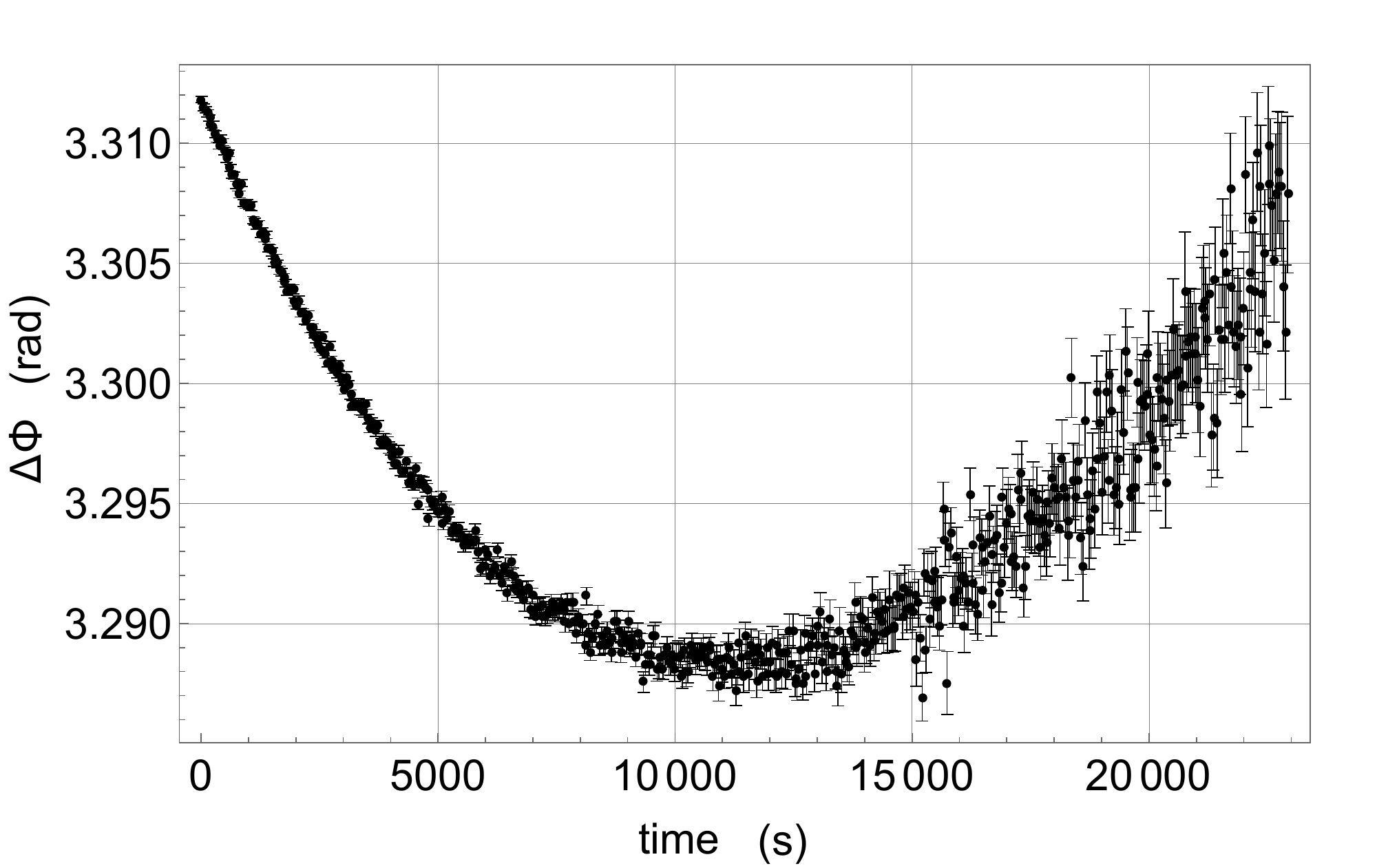}
\includegraphics[width=\columnwidth]{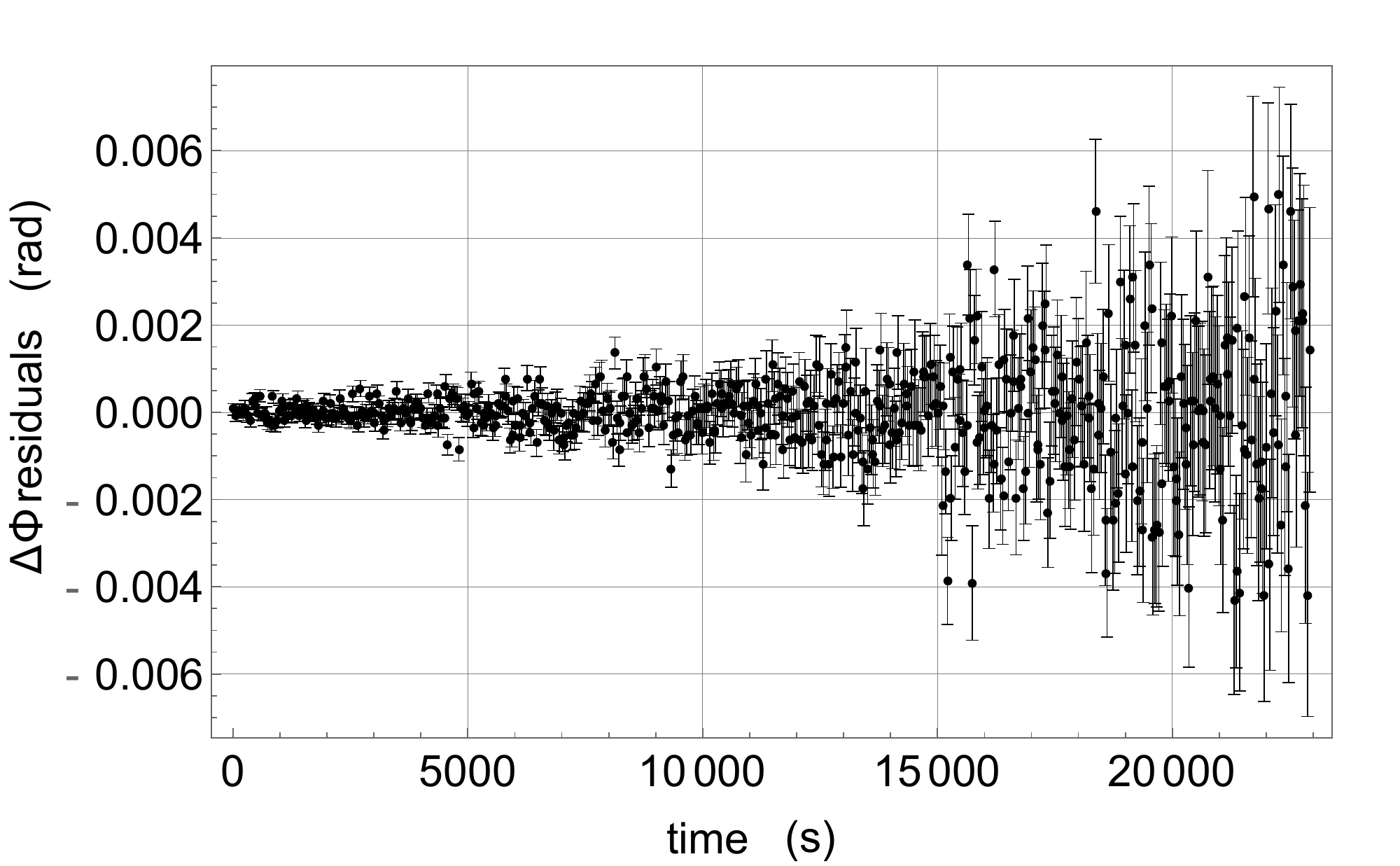}
\end{flushright}
\caption{Top: The weighted phase difference according to Eq.~(\ref{eqn:phasediff}) for measurement run number 6. Bottom: The weighted phase difference after subtraction of the fitted model in Eq.~(\ref{eqn:edmfit}) (residuals). Each data point corresponds to a time bin of 40~s.
\label{fig:phase}}
\end{figure}
\subsection{Deterministic phase shifts}
\label{sec:detphaseshifts}
Due to {\it Earth's rotation}, the laboratory-reference system is not an inertial frame. The SQUID detectors rotate with a frequency $\omega_\text{det}$ with respect to the precessing spins, and so, the measured precession frequencies of $^3$He and $^{129}$Xe are the actual Larmor frequencies shifted by $\omega_\text{det}$. In the weighted phase difference, this contribution is
\begin{eqnarray}
\Delta\Phi_\text{Earth}&=&\left(1-\frac{\gamma_\text{Xe}}{\gamma_\text{He}}\right)\omega_\text{det} t~~.
\end{eqnarray}
The sign and magnitude of $\omega_\text{det}$ depend on the orientation of the magnetic guiding field with respect to the Earth's rotation axis. For a magnetic guiding field in the horizontal plane at latitude $\Theta_1$ and angle $\rho_1$ to the north-south direction, this is
\begin{eqnarray}
\label{eqn:Earthrotation}
\omega_\text{det}&=&\Omega_E\cos(\rho_1)\cos(\Theta_1)~~,
\end{eqnarray}
where the sidereal frequency is given by $\Omega_E=2\pi/86164.101~\text{s}^{-1}$ and $\Theta_1=50.92^\circ$ (J\"ulich, Germany). The field of the cosine coil was at roughly pointed $\rho_1\approx45^\circ$ to the north-south direction. So, we expect: $\Delta\Phi_\text{Earth}\approx39\cdot10^{-6}\frac{\text{rad}}{\text{s}}\cdot t$. However, the exact orientation of the field is difficult to determine (uncertainty of about $1^\circ$). As a consequence, the actual contribution of Earth's rotation to the phase shift has to be determined by the fit to the weighted phase data.\\
Additional linear phase shifts arise due to the uncertainty of the ratio $\gamma_\text{Xe}/\gamma_\text{He}$ itself and {\it chemical shifts}. The most precise values available for the shielded $^3$He and $^{129}$Xe nuclear magnetic moments were derived from the ratio of Larmor frequencies of $^3$He and $^1$H~\cite{Flowers}, and $^{129}$Xe and $^1$H~\cite{Pfeffer}, respectively. Dividing the frequency ratios yields the quotient of interest for the EDM search in $^{129}$Xe:
\begin{eqnarray}
\frac{\omega_\text{Xe}}{\omega_\text{H}} /\frac{\omega_\text{He}}{\omega_\text{H}}=\frac{\gamma_\text{Xe}}{\gamma_\text{He}}&=&0.363097448(24)~~.
\label{eqn:gammaratios}
\end{eqnarray}
The ratio $\gamma_{Xe}/\gamma_{He}$ is valid  in the zero gas pressure limit, \textit{i.e.} free from intermolecular interactions. In \cite{Makulski}, the $^3$He and $^{129}$Xe frequencies (chemical shifts) were examined in gaseous mixtures of $^3$He/$^{129}$Xe, $^3$He/$^{129}$Xe/CO$_2$ and $^3$He/$^{129}$Xe/SF$_6$, \textit{i.e.}, the gas mixtures being used in our experiment. The density-dependent ($\rho_2$) chemical shift of $^{129}$Xe gives by far the strongest contribution with $\sigma_2=-13.506$~ppm$\cdot$liter/Mol. As a result of that, we obtain a linear shift in the weighted frequency difference given by:
\begin{eqnarray}
\Delta\omega_\text{CS}= \gamma_\text{Xe} B \left(-\sigma_2 \rho_2 \pm 24\cdot10^{-9}/(\gamma_\text{Xe}/\gamma_\text{He})\right)~~,
\label{eqn:chemicalshift}
\end{eqnarray}
where the second term accounts for the uncertainty in the ratio $\gamma_\text{Xe}/\gamma_\text{He}$ (Eq.~(\ref{eqn:gammaratios})). For Xe partial pressures of about 100~mbar ($\rho_2$= 0.00406 Mol/liter), the additional linear phase shift is $\Delta\Phi_\text{CS}= (-3.58~\text{to}~+0.32)\cdot10^{-6}\frac{\text{rad}}{\text{s}}\cdot t$, which is about an order of magnitude less than $\Delta\Phi_\text{Earth}$, but not negligible. 
The combined effects of chemical shift, the uncertainty of $\gamma_\text{Xe}/\gamma_\text{He}$ and Earth's rotation result in a phase shift that is linear in time: they are subsummed under $\Delta\omega_\text{lin}$ in the further 
course with $\Delta\omega_\text{lin}$ as a free fit parameter given in Eq.~(\ref{eqn:edmfit}) (below).\\
We consider a neutral particle (here, $^3$He or $^{129}$Xe) with spin and magnetic moment precessing steadily with the Larmor frequency $\omega_L$ in a constant magnetic field $B_z$. The addition of a rotating field with amplitude $B_1$ and frequency $\omega_D$ in the $x$-$y$-plane leads to a shift of the precession frequency, the {\it Ramsey-Bloch-Siegert (RBS) shift}:
\begin{equation}
\label{eqn:RBS}
\delta\omega_{\text{RBS}}(t)=\pm\left(\sqrt{\Delta\Omega^2+\gamma^2B_1^2}-\Delta\Omega\right)
\end{equation}
with $\Delta\Omega=|\omega_L-\omega_D|$. The plus sign applies to $\frac{\omega_D}{\omega_L}<1$, the minus sign to $\frac{\omega_D}{\omega_L}>1$, respectively. In our case, the rotating field is generated by the precessing magnetization of the $^3$He/$^{129}$Xe spin sample. Two effects contribute to the RBS shift: cross-talk (CT) and self-shift (SS). 
The cross-talk describes the shift due to the influence of the precessing magnetization of the $^{3}$He nuclei (with  $\omega_D=\omega_\text{He}$) on the $^{129}$Xe precession frequency (and vice-versa). Since $\gamma B_1\ll \Delta\Omega$ is fulfilled with $\Delta\Omega\approx 2\pi\cdot8$~Hz and $\gamma B_1<2\pi\cdot0.02$~Hz (estimated from the $B_1$ field of a uniformly magnetized sphere \cite{Jackson} which amounts  to $\approx600$ pT for our spin samples), the cross-talk results in a RBS frequency shift of
\begin{eqnarray}
\delta\omega_\text{CT}&\approx&\frac{\gamma^2B_1^2}{2\Delta\Omega}\,\,\,\,\,\,\,\,(\gamma B_1\ll\Delta\Omega)~~.
\end{eqnarray}
Thus, the shift in the $^{129}$Xe frequency due to the cross-talk is:
\begin{eqnarray}
\delta\omega_\text{CT,Xe}&=&\frac{\gamma_\text{Xe}^2 B_{1,\text{He}}^2}{2\Delta\Omega}~~.
\end{eqnarray}
In order to get the accumulated cross-talk phase, one integrates over time (omitting a constant term that can be absorbed into a final constant term in the description of all deterministic phase shifts) and inserts the exponential decay of the signal amplitude, \textit{i.e.} $B_{1,\text{He}}(t)=B_{1,\text{He}}(0) \exp\left(-t/T_{2, \text{He}}^*\right)$:
\begin{eqnarray}
\delta\Phi_\text{CT,\text{Xe}}(t)&=& -\frac{\gamma_\text{Xe}^2B_{1,\text{He}}(0)^2 T_{2, \text{He}}^*}{4\Delta\Omega} \exp\left(-\frac{2\cdot t}{T_{2, \text{He}}^*}\right)~.
\end{eqnarray}
A corresponding term can be derived for the CT phase shift on the precessing $^3$He magnetization. The time dependence is described by the two exponential terms with time constants $\frac{1}{2}T_{2, \text{Xe}}^*$ and $\frac{1}{2}T_{2, \text{He}}^*$. This is a direct result of the quadratic dependence on $B_1$. The amplitude has to be determined by the fit, since $B_{1,\text{He(Xe)}}(0)$ cannot be quantified with the required accuracy.\\
In contrast to the cross-talk, the self-shift occurs even when there is only one spin species present. The self-shift is a result of the coupling of the precessing magnetic moments of the same spin species among each other in the presence of an inhomogeneous magnetic field. The gradients of the magnetic guiding field $\vect{B_0}$ are in the order of 10~pT/cm, and we expect $\Delta \Omega \ll 10~\text{pT}\cdot \gamma$ including field averaging due to diffusion of the sample spins. Thus, the condition $\Delta \Omega \ll \gamma B_1$ is met and one derives for the self-shift from Eq.~(\ref{eqn:RBS}) in first-oder approximation:
\begin{eqnarray}
\label{eqn:SS}
\delta\omega_\text{SS}&=&\pm\gamma B_1~~.
\end{eqnarray}
The sign depends on the sign of $\omega_L-\omega_D$. In general, the self-shift strongly depends on the field gradients across the sample cell, the resulting diffusion coefficients for $^3$He and $^{129}$Xe in the gas mixture, and the shape of the sample cell~\cite{Gemmel}. However, during a single run, these parameters are sufficiently constant,  so that only the time dependence of $B_1(t)\propto \exp(-t/T_2^*)$ enters which results in the corresponding exponential behavior of the accumulated phase $\delta\Phi_\text{SS,He(Xe)}$, given by
\begin{eqnarray}
\label{eqn:SS2}
\delta\Phi_\text{SS,He(Xe)}&\propto&\exp(-t/T_2^*)~~.
\end{eqnarray}
The phase amplitude again has to be determined by the fit.
\subsection{Fit to weighted phase-difference data}
\begin{figure}
\begin{center}
\includegraphics[width=\columnwidth]{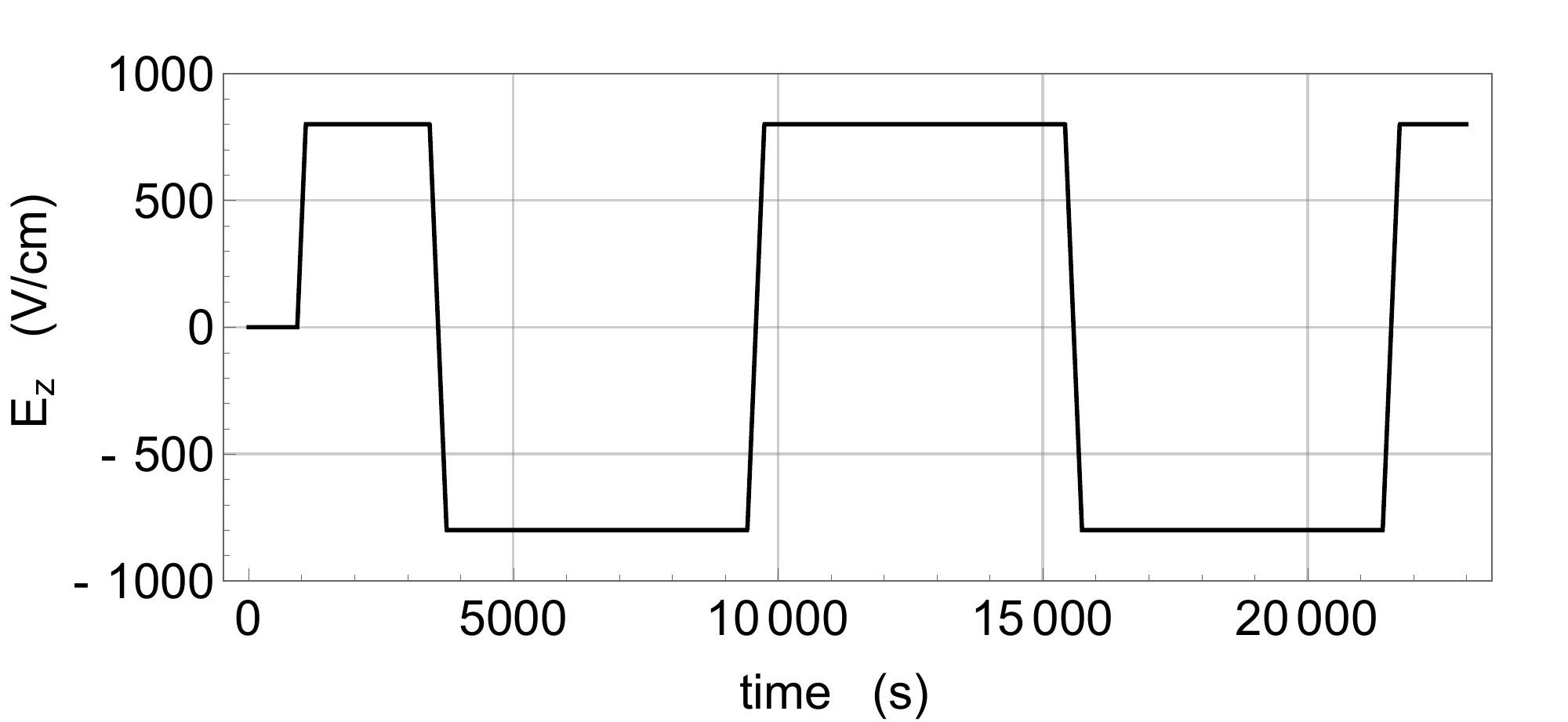}
\includegraphics[width=\columnwidth]{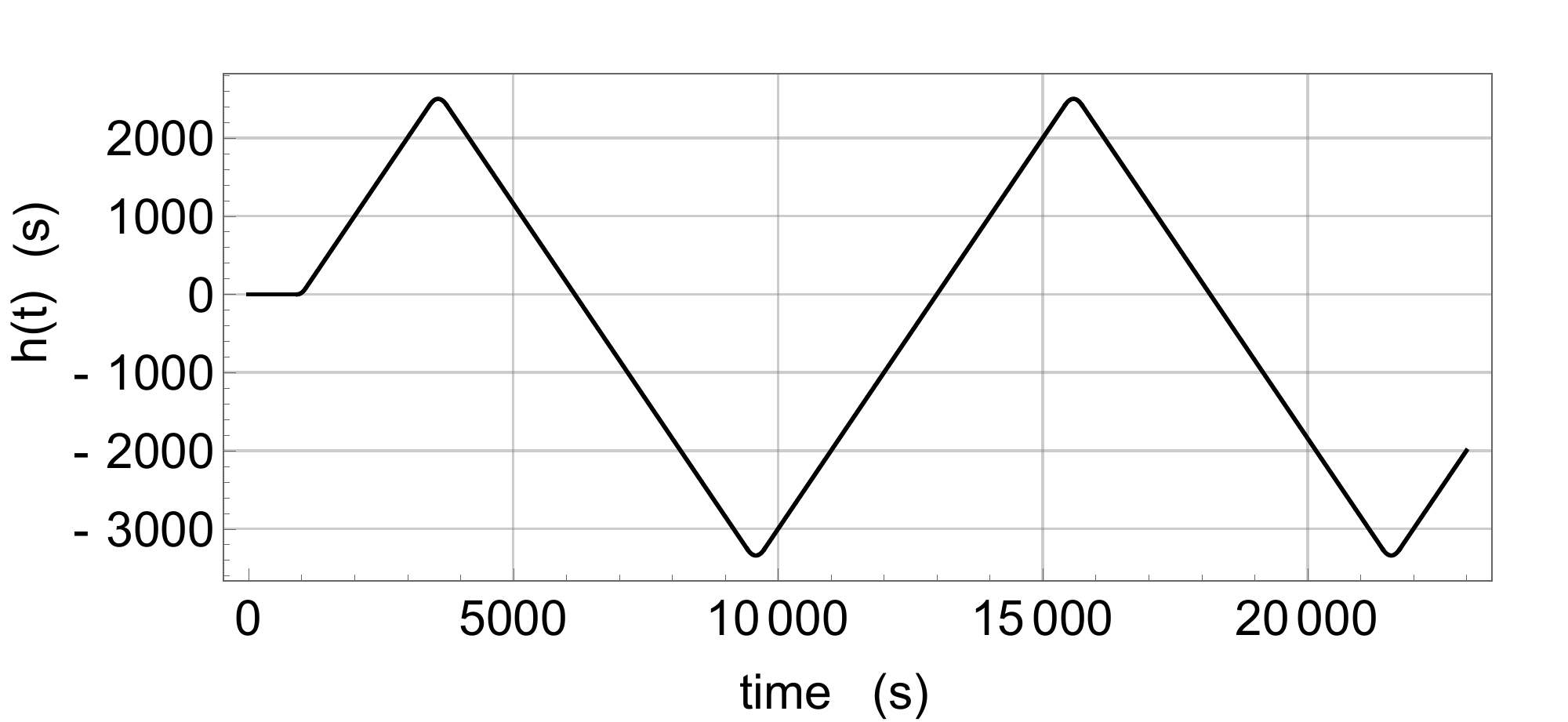}
\end{center}
\caption{An electric field that is periodically switched between $\pm E_{z}$ with a period $T_a=12000~$s in this example causes a signal in the weighted phase difference that is proportional to a triangular wave ($h(t,T_a)$) for a non-zero EDM. Due to the finite high voltage ramping rate of 25~V/s, the edges of the function are smoothed out.
\label{fig:EDMfunctions}}
\end{figure}

In the case of an electric field that is periodically switched between $\pm E_{z}$ as shown in Fig.~\ref{fig:EDMfunctions} (top), the weighted phase difference due to an EDM $d_\text{Xe}$ as given by Eq.~(\ref{eqn:frequencydiff2}) is proportional to the triangular wave $h(t,T_a)$ with period $T_a$ and slope of 1, resp. -1 (see Fig. \ref{fig:EDMfunctions} (bottom)):
\begin{eqnarray}
\label{eqn:deltaphiedm}
\Delta\Phi_\text{EDM}(t)&=&\frac{2d_\text{Xe}}{\hbar}E_z\cdot h(t,T_a):=g\cdot h(t,T_a)~~.
\end{eqnarray}
The appropriate fit function to the weighted phase-difference data includes all deterministic phase shifts, a trivial phase offset, and the parametrization of an EDM-induced phase shift $g\cdot h(t,T_a)$. It is given by
\begin{eqnarray}
\label{eqn:edmfit}
&&\nonumber\Delta\Phi_{\text{fit}}\left(t\right)=\\
\nonumber
&&\Phi_{0}+\Delta\omega_{\text{lin}} t +E_\text{He}\exp\left(\frac{-t}{T_{2,\text{He}}^*}\right)+E_\text{Xe}\exp\left(\frac{-t}{T_{2,\text{Xe}}^*}\right)\\
&&+F_\text{Xe}\exp\left(\frac{-2t}{T_{2,\text{He}}^*}\right)+F_\text{He}\exp\left(\frac{-2t}{T_{2,\text{Xe}}^*}\right)+g h(t,T_a)
\end{eqnarray}
As the correlation of fit parameters ($\Phi_{0}$, $\Delta\omega_{\text{lin}}$, $E_\text{He(Xe)}$, $F_\text{He(Xe)}$ and $g$) can be very high in particular cases, several fitting algorithms were compared to validate the results. An analytical and a purely numerical least square fitter, as well as a maximum likelihood fitter were tested. It proved useful to orthogonalize the individual terms of the fit model to reduce correlation and thereby increase numerical stability. In Appendix \ref{sec:ortho}, details of the orthogonalization procedure can be found. The different algorithms returned the same fit results (within numerical noise). Finally, estimations for the fit-parameter values including their correlated and uncorrelated uncertainties are extracted. Additionally, the reduced $\chi^2$ as a measure of the goodness of the fit, and the correlations of the fit parameters are evaluated. In the next step, the atomic EDM of $^{129}$Xe is calculated from the fit parameter $g$:
\begin{eqnarray}
d_\text{Xe}&=&\frac{\hbar}{2 E_{z}} g~.
\end{eqnarray}
The corresponding uncorrelated and total (combination of uncorrelated and correlated) uncertainties $\delta d$ are determined for a separate test run (no applied electric field) with total data acquisition time of seven hours. The data of this run are analyzed for a set of different (simulated) $E$-field switching periods $T_a$ in order to find the highest EDM sensitivities (see Fig. \ref{fig:EDMerrorvsta}):
\begin{figure}
\includegraphics[width=\columnwidth]{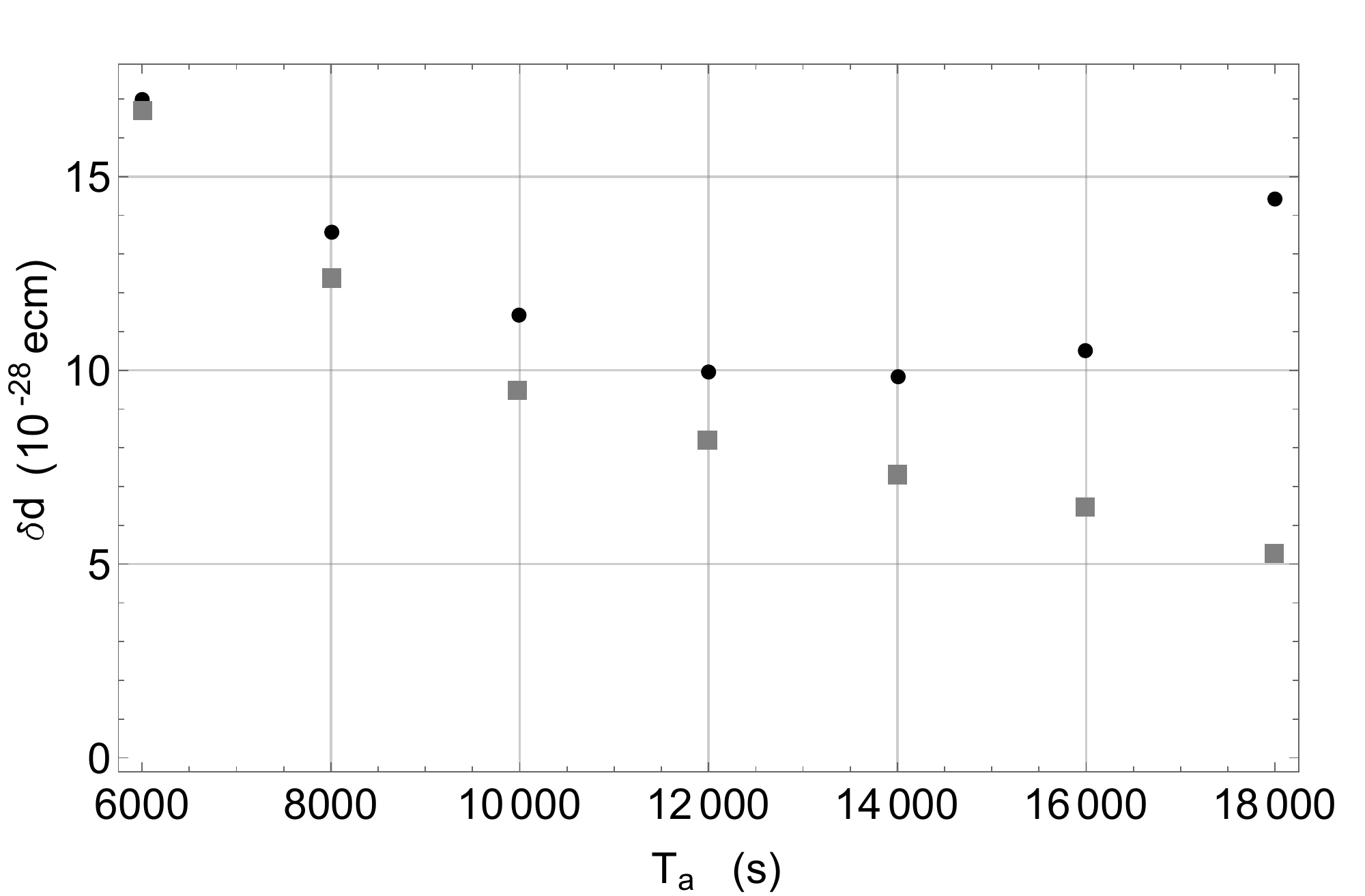}
\caption{The resulting total (black circles) and uncorrelated (gray squares) EDM uncertainty of a separate test run (without physical $E$-field switching but assuming a hypothetical field of $E_z= 800$~V/cm lasting approximately seven hours with $T_\text{2, Xe}^*=2.9$~h as a function of the $E$-field switching period $T_a$.
\label{fig:EDMerrorvsta}}
\end{figure}
in general, it is found that the uncorrelated uncertainty of the EDM $d$ decreases with larger $T_a$. For short $T_a$, the contribution of the correlated uncertainty to the total uncertainty is small because the correlation between $h(t,T_a)$ and the other time-dependent terms describing the deterministic phase shifts is very small. However, with larger $T_a$, the correlation increases (especially with the exponential terms describing the Ramsey-Bloch-Siegert shift), resulting in correlated uncertainties that are a factor of $\approx 3$ higher than the uncorrelated one, \textit{e.g.} by choosing $T_a=18000$~s.
For the analyzed test run with no applied electric field, a relatively flat optimum is found at $T_a\approx 14000$~s (see Fig.~\ref{fig:EDMerrorvsta}). The total EDM sensitivity in this case (assuming a hypothetical field of $E_z= 800$~V/cm) reaches $\delta d=10^{-27}$~\textit{e}cm.
\subsection{Comparison of cylindrical and spherical EDM cells}
The use of EDM cells of spherical symmetry has its reasons in the suppression of phase drifts caused by the RBS self-shift with unknown time structure. Spin-probes of spherical symmetry do not show demagnetization effects, which produce sample inherent gradients across the cell volume. For a cylindrical cell (diameter 10~cm, length 5~cm) with a magnetization of 600~pT/$\mu_0$, these field gradients can reach 50~pT/cm \cite{Caciagli}. During spin-precession, the rotating transverse magnetization leads to rotating gradients; and if there is a finite longitudinal magnetization left (imperfect spin flip), to an additional spatially static gradient. In both cases, these demagnetization induced gradients decrease with time as the magnetization of the spin sample relaxes towards zero. For the resulting RBS self shift, this implies that its time behavior can no longer be described by a simple exponential term (see Eq.~(\ref{eqn:SS2})), since the prefactor now becomes time-dependent, too, and cannot be parameterized with the required accuracy. Therefore, the concept of cylindrical EDM cells with lid electrodes made of silicon, which was originally approached, was discarded in favor of spherical sample cells. The drawback with external electrodes in the form of a plate capacitor is that one has to guarantee that the electric field inside the insulating glass cell is essentially the externally applied field. To avoid Townsend-type gas discharges which may compensate the external electric field, only a moderate electric field of 800~V/cm was applied. The electric field and its temporal behavior inside the glass bulb filled with the same gas mixture as in the experiment was investigated in extensive off-line tests by means of an electro-optic field sensor based on a LiNbO$_3$ crystal~\cite{Grasdijk}; and an electro-mechanical field-mill sensor, as well as on-line by the pA-meters (see Appendix~\ref{sec:efieldmeasurement}). A limit for a possible decay of the field amplitude was deduced. It could be concluded that inside the EDM glass cell the electric field strength was on the average larger than 95\% of the externally applied field. Therefore, when extracting our Xe-EDM limit from our data, we must replace $E_z$ with $E_{z,\text{eff}}=0.95\cdot E_z$ at the respective places.
\subsection{Leakage current measurements}
\begin{figure}
\includegraphics[width=\columnwidth]{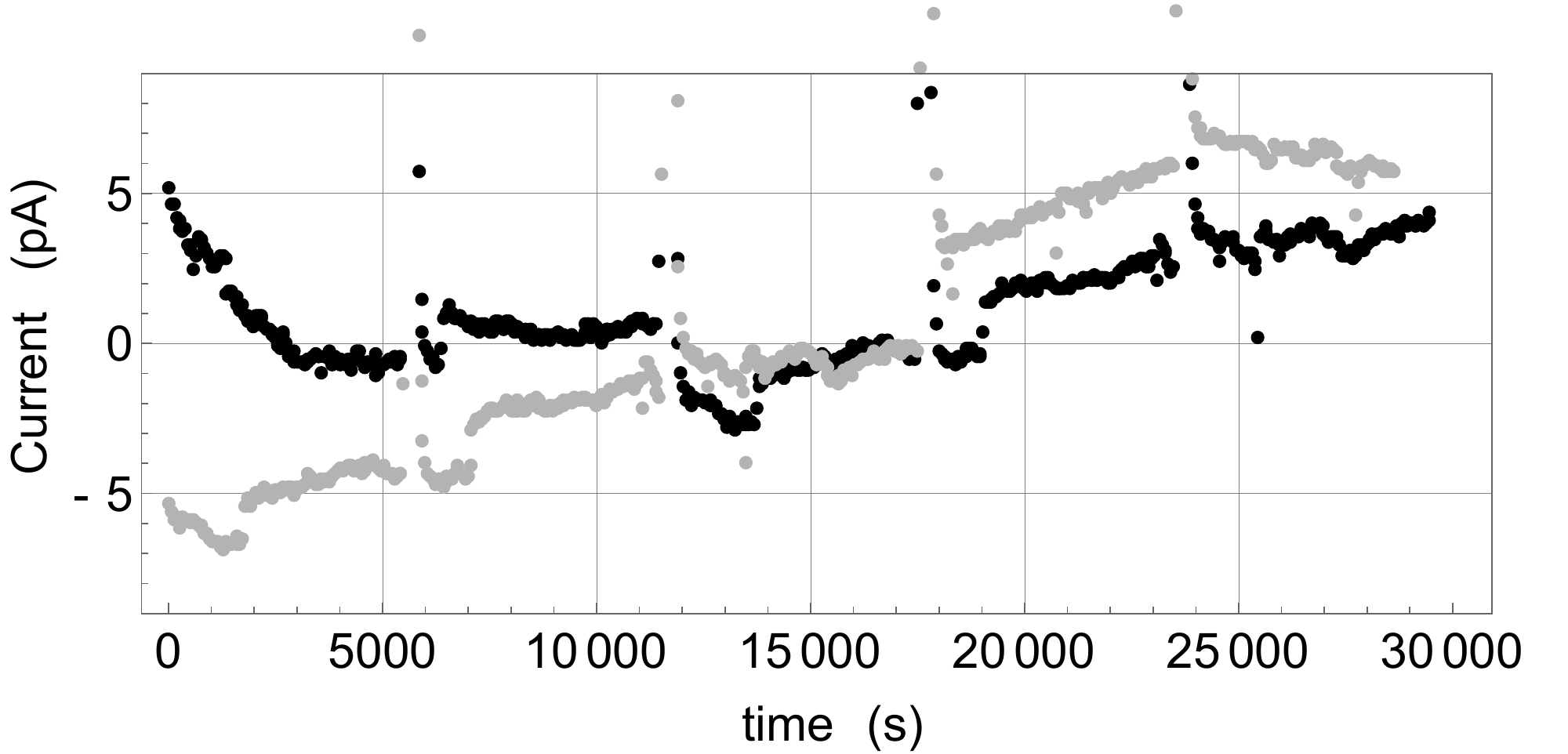}
\caption{Leakage currents monitored during EDM run number 6 with the two pA-meters in the high voltage supply lines to the electrodes (see Fig.~\ref{fig:cell}). The pA-meters have offsets that slightly drift with time. During polarity reversal a displacement current of $\approx40$~pA is detected (data points are truncated for a better presentability. In Fig.~\ref{fig:pAdata2} the current during field switching is displayed). Error bars are smaller than the symbol size.
\label{fig:pAdata}}
\end{figure}

In considering magnetic systematic effects in EDM measurements, the prime suspect is always leakage currents. Any currents flowing near the EDM cell during the recording phase of the coherent spin precession can generate magnetic fields that directly lead to HV-correlated phase shifts. For example, a helical current path along the walls of the  EDM cell between the oppositely charged electrodes would create a magnetic field component that adds linearly to the magnetic holding field $B_0$,  producing a Larmor-frequency shift with the same $E$-field dependence as an EDM. Ideally, co-magnetometry compensates such leakage current-induced effects. Higher order effects, however, may cause leakage current-induced EDM false effects as discussed in Section~\ref{sec:syseffects}. From that point of view, it is still of utmost importance to have a sensitive monitoring of the leakage currents. During the EDM-measurement runs, the leakage currents are constantly monitored by the pA-meters depicted in Fig.~\ref{fig:cell} and do not exceed a few pA at an applied electric field of 800~V/cm (see Fig.~\ref{fig:pAdata}). As the direction of the electric field has to be inverted repeatedly over the course of an individual run, displacement currents flow to charge and discharge the electrodes that have a capacitance of about 1.5~pF.
\subsection{Results}
\begin{figure}
\includegraphics[width=\columnwidth]{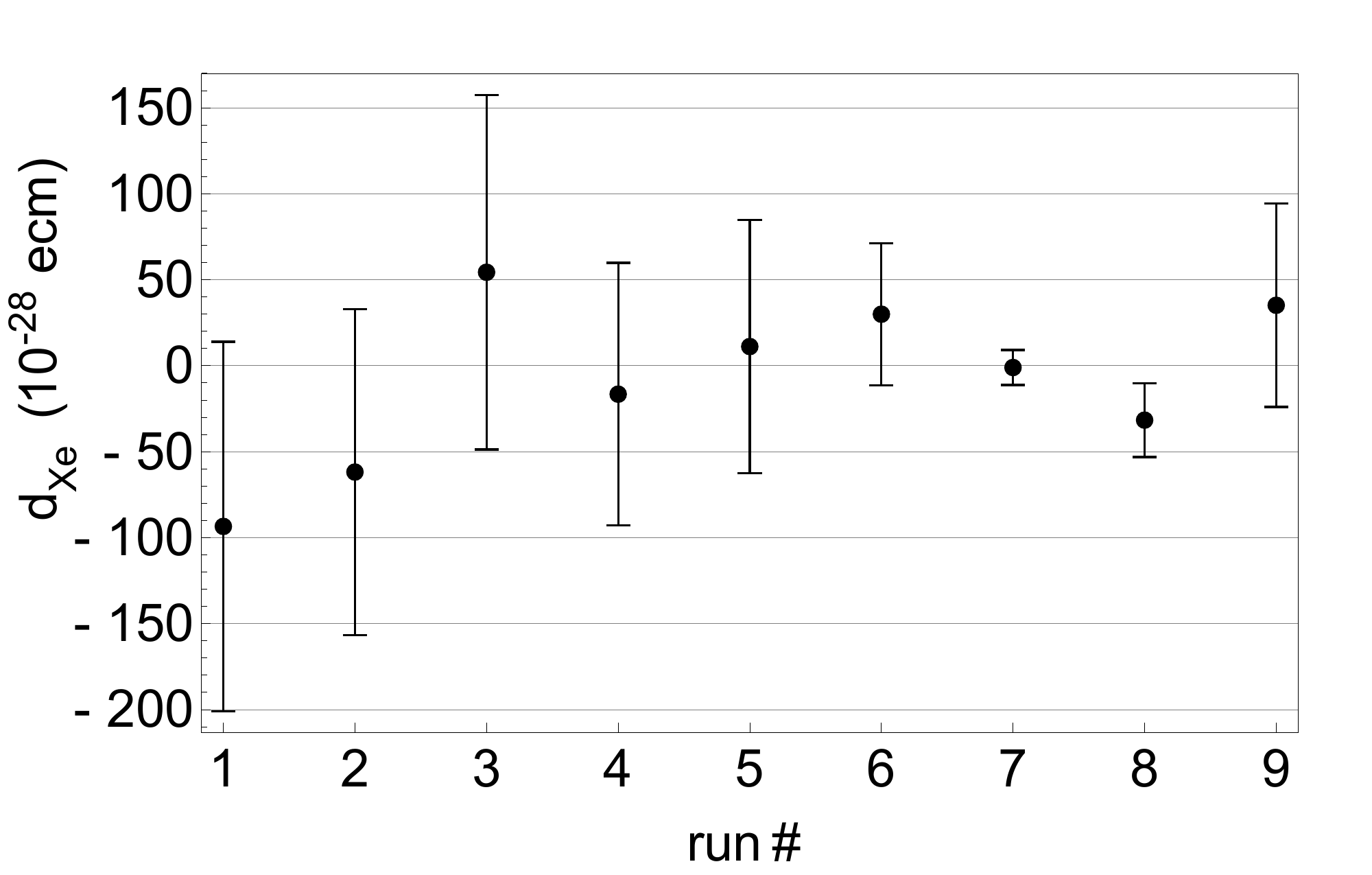}
\caption{The extracted EDM mean values and statistical uncertainties (including correlated uncertainties) of the nine single runs. The big difference in the resulting total uncertainty of the single measurement runs (lasting between 5 and 11 hours) is striking. The reason is the strong dependence of the statistical uncertainty (Eq.~(\ref{eqn:sensitivity})) of a single EDM run on the parameter settings (see Tab.~\ref{tab:result}).
\label{fig:results}}
\end{figure}

Nine independent runs have been performed where the partial pressures of hyperpolarized $^3$He, $^{129}$Xe and the buffer gases SF$_6$ and CO$_2$ were varied. The period of the electric field reversal $T_a$ was adjusted to the relaxation time of xenon and varied between 12000~s and 18000~s. The typical length of a single run was between 5 and 11~h. The starting polarity (magnetic and electric field either parallel or anti-parallel) was varied. We analyzed the data first without considering the sign of the initial polarity and only took the real polarity with the corresponding sequence of the electric field reversal according to Fig.~\ref{fig:EDMfunctions} into account in the last step (unblinding). No significant difference between the two polarity groups was found. Table~\ref{tab:result} summarizes the relevant parameters (partial pressures, measured relaxation times, etc.) of the individual runs. 
The extracted EDM limits of the individual runs are shown in Fig.~\ref{fig:results}. One finds that the statistical uncertainty decreased considerably (by a factor of 10 between run number 1 and 7) due to a successive optimization of the experimental parameters. These are mainly the partial pressures of $^3$He, $^{129}$Xe and the buffer gases in order to reach a high SNR in combination with long spin coherence times; but, also the steady improvement of the Xe polarization.\\
The combined mean value of nine $^{129}$Xe-EDM measurement runs (derived from a combined fit that reduces correlated uncertainties) is
\begin{eqnarray}
d_\text{Xe}&=&\left(-4.7\pm6.4\right)\cdot 10^{-28}~e\text{cm~~.}
\end{eqnarray}
\begin{table*}
\begin{tabular}{l|rrrr|rr|rr|r|c|r|rl}
\hline
\hline
\# &$p_\text{He}$&$p_\text{Xe}$&$p_\text{SF6}$&$p_\text{CO2}$&$A_{\text{He}}$/pT&$T_{2,\text{He}}^*$/h&$A_{\text{Xe}}$/pT&$T_{2,\text{Xe}}^*$/h&$T$/h&$T_a$/s&red.~$\chi^2$&\multicolumn{2}{|c}{$d~/~10^{-28}$~\textit{e}cm}\\
&\multicolumn{4}{|c|}{mbar}&&&&&&&&\\
\hline
1&38&20&5&20&17.8&3.0&11.8&1.8&5.8&$1.2\cdot10^4$&1.25&-93.5&$\pm$107.4\\
2&22&18&3&20&5.8&4.4&8.7&2.4&7.2&$1.2\cdot10^4$&1.18&-61.9&$\pm$94.7\\	
3&12&42&3&21&11.8&3.6&19.6&1.7&5.6&$1.2\cdot10^4$&2.60&54.3&$\pm$103.1\\	
4&12&24&4&49$^{*)}$&30.2&4.7&21.8&2.1&6.4&$1.6\cdot10^4$&1.45&-16.6&$\pm$76.4\\
5&25&53&5&44$^{*)}$&59.1&3.2&52.6&1.6&5.0&$1.2\cdot10^4$&1.26&11.2&$\pm$73.5\\	
6&45&96&0&0&128.4&18.9&113.4&2.9&6.5&$1.2\cdot10^4$&0.85&30.0&$\pm$41.3\\	
7&20&100&0&0&77.4&20.8&101.8&2.8&11.2&$1.8\cdot10^4$&1.33&-1.1&$\pm$10.2\\	
8&27&91&0&0&96.2&20.0&123.5&2.9&9.7&$1.8\cdot10^4$&1.40&31.7&$\pm$21.5\\	
9&31&103&0&0&104.7&18.0&117.1&2.8&6.6&$1.8\cdot10^4$&1.28&-35.1&$\pm$59.1\\	
\hline
\hline

\end{tabular}
\caption{
Compilation of the measurement parameters met in the individual EDM runs. In the sequence of columns: a) partial pressures of He, Xe and buffer gases, b) initial signal amplitude $A_\text{He(Xe)}$ and transverse relaxation time $T_{2,\text{He(Xe)}}^*$ for helium (xenon), c) total data acquisition time $T$, d) electric field switching period $T_a$, e) reduced $\chi^2$ of the fit, and f) extracted EDM value $d_\text{Xe}$.\\
 $^{*)}$ Here, $^4$He instead of CO$_2$ was used.
\label{tab:result}}
\end{table*}
\section{Phase stability and evaluation of noise and sensitivity}
\begin{figure}
\includegraphics[width=\columnwidth]{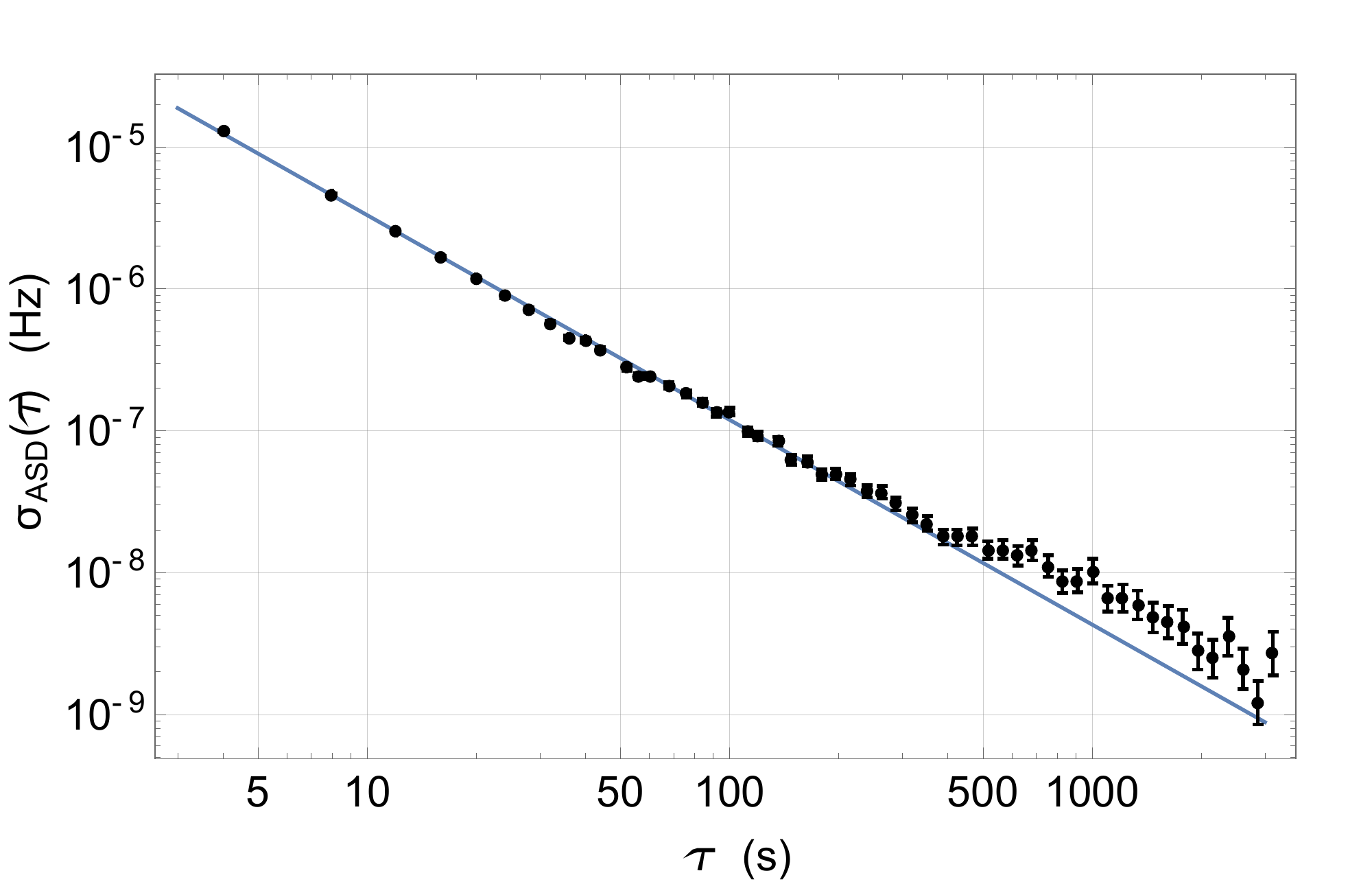}
\caption{ASD of the residual frequency noise which decreases with $\sigma_\text{ASD}\propto \tau^{-\frac{3}{2}}$ as indicated by an according fit (blue). This shows the presence of white noise (with small deviations around $\tau\approx1000$~s). To fulfill the ASD statistics criteria $N\gg 1$~\cite{Lesage}, only data are shown for integration times $\tau<3000$~s.
\label{fig:ASD}}
\end{figure}
The Allan Standard Deviation (ASD) \cite{Allan,Allan2, Barnes} is the most convenient measure to study the temporal characteristics of the $^{3}$He-$^{129}$Xe co-magnetometer and to identify the power-law model for the frequency and phase-noise spectrum. Deviations from the CRLB power law (Eq.~(\ref{eqn:sensitivity})) due to non-Gaussian noise sources can be traced by this data-analysis tool.
The ASD of the phase residuals (after subtraction of all deterministic phase shifts) is calculated according to:
\begin{eqnarray}
\sigma_\text{ASD}^\Phi(\tau)&=&\sqrt{\frac{1}{2N-2}\sum_{j=1}^{N}\left[\overline{\Delta\Phi}_{j+1}(\tau)-\overline{\Delta\Phi}_{j}(\tau)\right]^2}~~,
\end{eqnarray}
where the total acquisition time $T$ is subdivided into $N$ smaller time intervals of equal length $\tau$, so that $N\tau=T$. For each of such a sub-dataset ($j=1,2,...,N$), the mean of the phase residuals $\overline{\Delta\Phi}_{j}(\tau)$ is determined. For white Gaussian noise (one essential assumption the CRLB is based on), $\sigma_\text{ASD}^\Phi$ coincides with the classical standard deviation and we expect $\sigma_\text{ASD}^\Phi\propto\tau^{-1/2}$.\\
The corresponding ASD of the frequency $f$ is calculated by dividing $\sigma_\text{ASD}^\Phi(\tau)$ by $2\pi\tau$. The frequency ASD ($\sigma_\text{ASD}$) plot for the phase residuals of run number 6 is shown in Fig.~\ref{fig:ASD}. With increasing integration times $\tau$, the uncertainty in frequency decreases down to the nHz level. The ASD plot shows the $\sigma_\text{ASD}\propto \tau^{-\frac{3}{2}}$ behavior according to the CRLB in Eq.~(\ref{eqn:sensitivity}), with slight deviations for  integration times $\tau\approx1000$~s. This behavior results in increased reduced $\chi^2$ values of the fit (see Tab.~\ref{tab:result}). Correspondingly, the statistical uncertainties of the fit-parameters (including $g$ for the extraction of $d_\text{Xe}$) were scaled with $\sqrt{\chi^2/\nu}$.

\section{Potential systematic effects}
\label{sec:syseffects}
Here, we discuss mechanisms which might generate a signal with the same signature as an EDM when an electric field is applied. Understanding and limiting the size of potential systematic effects is an extremely important part of performing a high-precision EDM measurement. A systematic effect would have to cause a shift in the $^{129}$Xe spin-precession frequency that is correlated with the applied HV polarity. While generating a false EDM signature, it is also possible that a systematic effect could cancel a signal from a real EDM, and thus, giving a false null measurement. In the following, we only discuss systematic effects that might lead to false EDM signals larger than $10^{-33}~e$cm. 
\subsection{HV-correlated magnetic field gradients}
Possible sources of high-voltage correlated magnetic fields are leakage currents or the displacement current during polarity reversal of the electric field.
From Fig.~\ref{fig:pAdata}, it can be safely deduced that leakage currents do not exceed a few pA at an applied electric field of 800 V/cm. If an assumed electric-field correlated leakage current of $I=5$~pA flows in a circular loop of $R=5$~cm (radius of the EDM cell) between the electrodes (which is indeed a worst-case scenario), then the maximum field gradients reach $1\cdot 10^{-17}$~T/cm. Reversing the polarity of the electric field leads to a displacement current of $I\approx50$~pA. Although 10 times higher than the assumed leakage current, its  time average leads to smaller effective field gradients.\\
In principle, the effects of HV-correlated magnetic fields should be eliminated by co-magnetometry via the analysis method of the weighted frequency or phase difference (\textit{cf.}~Eqs.~(\ref{eqn:frequencydiff}) and (\ref{eqn:phasediff})). However, two residual effects can be identified which are attributed to the gradients of such fields:\\
firstly, the difference in the molar masses of $^3$He and $^{129}$Xe leads to a difference in their centers of masses (barometric formula), which is $\Delta \overline{y}=0.31~\mu\text{m}$ for our spherical sample cell. A gradient along the vertical axis $\partial B / \partial y$ causes a non-vanishing weighted frequency difference of $\Delta \omega_\text{grav}=\gamma_\text{Xe}\Delta \overline{y}\partial B / \partial y$. The corresponding false EDM due to this {\it gravitational shift} is
\begin{eqnarray}
d_\text{grav}&=&\gamma_\text{Xe}\Delta \overline{y}\frac{\partial B}{ \partial y}\frac{\hbar}{2E_z}~~.
\label{eqn:dgrav}
\end{eqnarray}
Therefore, leakage-current induced gradients of $1\cdot 10^{-17}$~T/cm give a maximum false EDM signal of $d_\text{grav}=8.5\cdot 10^{-33}$~\textit{e}cm. One might wonder that in the presence of field gradients, a spatial frequency dependence may arise since the SQUID receives more signal from the precessing spins that are close to the sensor than those further away. However, we can exclude such an effect, as the measured accumulated phase represents an excellent volume average due to spin diffusion (rapid sampling of the cell volume), and, furthermore, any residual effect will be the same for He and Xe and therefore drops out by comagnetometry.\\
Secondly, magnetic field gradients influence the transverse relaxation times $T_2^*$. Analytical expressions can be derived for spherical sample cells, as reported in \cite{Cates}: 
\begin{eqnarray}
\frac{1}{T_2^*}&=&\frac{1}{T_1}+\frac{8R^4\gamma^2}{175D}\cdot\left(\frac{\partial B}{\partial z}\right)^2~~.
\label{eqn:T2*}
\end{eqnarray}
Here, $R$ is the radius of the EDM cell, $D$ the diffusion coefficient in the gas mixture, and $T_1$ is the longitudinal relaxation time. The gradients $\partial B / \partial z$ are the superposition of gradients resulting from ambient influences $(\partial B / \partial z)_0$ and magnetic field gradients $(\partial B / \partial z)_\text{sys}$ that are correlated with the high voltage reversal. Since $(\partial B / \partial z)_\text{sys}\ll (\partial B / \partial z)_0$, the change in $T_2^*$ is:
\begin{eqnarray}
\Delta T_2^*&=&\left(T_2^*\right)^2\frac{16R^4\gamma^2}{175D}\left(\partial B / \partial z\right)_0 (\partial B / \partial z)_\text{sys}~~.
\label{eqn:DeltaT2*}
\end{eqnarray}
Under typical operating conditions and using the conservatively estimated maximum field gradients of $(\partial B / \partial z)_\text{sys}=1\cdot 10^{-17}$~T/cm, one finds that $\Delta T_{2,\text{He}}^*\approx 0.15$~s and $\Delta T_{2,\text{Xe}}^*\approx 0.005$~s. In the fit model describing the weighted phase difference data (see Eq.~(\ref{eqn:edmfit})), the change in $T_2^*$ leads to additional HV-correlated (almost) linear drifts of the weighted phase difference (as $\Delta T_2^*\ll T_2^*$):
\begin{eqnarray}
\exp\left(\frac{-t}{T_2^*+\Delta T_2^*}\right)\approx\exp\left(\frac{-t}{T_2^*}\right)\left(1+\frac{\Delta T_2^*}{\left(T_2^*\right)^2}t\right)~.
\end{eqnarray}
Those terms are highly correlated with the triangular term describing the EDM effect (Eq.~(\ref{eqn:deltaphiedm})) and give a false EDM signal of:
\begin{eqnarray}
d_\text{T2,He(Xe)}&=&\frac{\hbar}{2E_z}E_\text{He(Xe)}\frac{\Delta T_{2,\text{He(Xe)}}^*}{\left(T_{2,\text{He(Xe)}}^*\right)^2}~~.
\label{eqn:dT2}
\end{eqnarray}
Only helium contributes substantially to this effect. With $ T_{2,\text{He}}^*\approx$20~h, a self shift amplitude of $E_\text{He}=$0.8~rad (result of the fit according to Eq.~(\ref{eqn:edmfit})), and $E_z=800$~V/cm, one finds: $d_\text{T2}=8.5\cdot10^{-30}$~\textit{e}cm.

\subsection{Motional magnetic field}
An atom moving with velocity $\vect{v}$ through a region of non-zero electric field experiences a magnetic field
\begin{eqnarray}
\label{eqn:motionalmagfield0}
\vect{B_m}&=&\frac{1}{c^2} \vect{E}\times\vect{v}~~(\text{for}~v\ll c)
\end{eqnarray}
in its rest frame (where $c$ is the speed of light). If the angle $\Theta_{EB}$ between the electric field and the laboratory magnetic field $\vect{B_0}$ is small, the magnitude of the effective magnetic field experienced by the atoms is:
\begin{eqnarray}
\label{eqn:motionalmagfield}
B&=& B_0+\frac{\Theta_{EB}  v_{\perp}  E_z}{c^2}+\frac{v_{xy}^2E_z^2}{2 c^4 B_0}~~,(B_m\ll B_0)~~.
\end{eqnarray}
Here, $v_{\perp}$ is the component of $v$ that is perpendicular to the plane of $E$ and $B$, and $v_{xy}=\sqrt{v_x^2+v_y^2}$.
$\vect{B_m}$ can lead to an EDM-like systematic shift under two conditions: first, if $\Theta_{EB}\neq 0$, the precession frequency can shift linearly with the electric field strength, and second, even with $\Theta_{EB}= 0$, $\vect{B_m}$ can produce a false EDM if the electric field magnitude changes when the polarity is reversed.\\
In storage experiments (as in case of the $^3$He/$^{129}$Xe co-magnetometer setup), the {\it linear term} is suppressed in first order. Finite $\vect{E}\times\vect{v}$ shifts, however, can still arise if the average velocity $\langle\vect{v}\rangle$ for the polarized $^3$He and $^{129}$Xe atoms is non-zero. Such a case may occur, for example, if the spins preferentially relax at a single point on the wall of the EDM cell. To place an upper limit on this effect, one can estimate how the distribution of polarized atoms evolve under the influence of this relaxation source (similar to the discussion in \cite{Swallows}). To determine the magnitude of this translation, we consider the one-dimensional diffusion equation of the polarization $P$:
\begin{eqnarray}
\label{eqn:diffusion}
\frac{\text{d}P}{\text{d}t}&=& D \frac{\partial^2P}{\partial x^2}
\end{eqnarray}
in the range of $[-R,R]$, where the center (of the cell) is at $x=0$ and the single point-like source of relaxation is at $x=-R$. Further, the diffusion constants are $D_\text{He} \approx 6.02~\text{cm}^2$/s and $D_\text{Xe}\approx0.62~\text{cm}^2$/s at the experimentally relevant gas pressures. The general solution taking into account only the first two diffusion modes is:
\begin{eqnarray}
\label{eqn:diffusion2}
\nonumber
P(x,t)&\propto&\sqrt{T_2^* D} \cos\left(\frac{x-R}{\sqrt{T_2^* D}}\right)\exp\left(-\frac{t}{T_2^*}\right)\\
&&+\frac{2R}{\pi} \cos\left(\frac{\pi (x-R)}{2R}\right) \exp\left(-\frac{\pi^2 D t}{4R^2}\right)~,
\end{eqnarray}
where the respective transverse relaxation times are $T_{2,\text{He}}^*$=20~h and  $T_{2,\text{Xe}}^*$=2.9~h. The second term decays very fast with time constants of 1.7~s (helium) and 16~s (xenon) and can be neglected in the further course, as the steady-state is reached long before the electric field is applied (after $\approx300$~s).
The polarization-weighted mean velocity can then be expressed as
\begin{eqnarray}
\label{eqn:diffusion3}
\langle v(t) \rangle&=&\frac{ -D\int_{-R}^{R}  P(x,t) \frac{\text{d}P(x,t)}{\text{d}x}~\text{d}x}{\int_{-R}^{R}  P(x,t)~\text{d}x}~~.
\end{eqnarray}
For $^3$He and $^{129}$Xe, we finally obtain (with $R$=5~cm):
\begin{eqnarray}
\label{eqn:diffusion4}
\langle v(t=0) \rangle_\text{He}&=&-4.6~\mu\text{m/s}\\
\langle v(t=0) \rangle_\text{Xe}&=&-3.9~\mu\text{m/s}
\end{eqnarray}
at the beginning of the measurement (where the effect is maximal). Realistically, one would have to compute the (weighted) average over the period $T_a/2$. However, we took $\langle v(t=0) \rangle$ as a conservative estimation. The ensemble average of the frequency shift (linear term of Eq.~(\ref{eqn:motionalmagfield}), $v_{\perp}$ replaced by $\langle v \rangle$ ) is
\begin{eqnarray}
\label{eqn:omegam}
\langle\delta \omega_m\rangle&=&\gamma \langle v\rangle E_z \Theta_{EB} /c^2
\end{eqnarray}
which by use of Eqs.~(\ref{eqn:elevel}) and (\ref{eqn:frequencydiff}) gives rise to a false EDM of
\begin{eqnarray}
\label{eqn:dm}
d_m&=&\frac{\hbar}{2E_z}\langle\Delta \omega_m\rangle=\frac{\hbar\gamma_\text{Xe}\left(\langle v\rangle_\text{Xe}-\langle v\rangle_\text{He}\right) \Theta_{EB}}{2c^2}~~.
\end{eqnarray}
Assuming (very conservatively) that $\Theta_{EB} < 0.03$~rad, this gives a false EDM of
\begin{eqnarray}
\label{eqn:diffusion7}
d_m = 5.8\cdot10^{-31}~e\text{cm}~~.
\end{eqnarray}
It should be stated that assuming a single point of relaxation inside the EDM cell is overly pessimistic since there are generally many tiny magnetic sites distributed on the surface of the glass vessel. This is discussed, \textit{e.g.}, in \cite{Schmiedeskamp}. Therefore, the linear motional magnetic field effect is much smaller in reality. We performed finite element simulations (using Comsol and Mathematica) to determine electric field homogeneity, \textit{i.e.} the (position dependent) angle $\Theta_{EB}$, considering various imperfections like misalignment of cell and electrodes, inhomogeneous wall thickness, etc., and found an volume average $\Theta_{EB} < 0.02$. The contribution of electric field inhomogeneity to the motional magnetic field effect is smaller than the conservative estimate of Eq.~(\ref{eqn:diffusion7}).\\
In \cite{Swallows}, the effect of convection inside an EDM cell, which may lead to additional motional magnetic field effects, was investigated. As result, an upper limit on the convection systematic error of $d_\text{conv} =2.2\cdot 10^{-31}$~\textit{e}cm was derived, assuming that a local heat source with a power of $0.3~\mu$W is deposited into the sample cell. In our case, no heat sources like lasers are used to monitor the spin-precession signal. Furthermore, the whole EDM cell is within a casing (T-shaped glass tube)  filled with SF$_6$ which thermally stabilizes the whole sample volume and keeps temperature differences across the cell much below 1~K. Therefore, we expect that one can safely use the estimate on $d_\text{conv}$ as an upper limit.\\
To determine the effects of the {\it quadratic term} in Eq.~(\ref{eqn:motionalmagfield}), one must consider the stochastic movement of the gas particles in the measurement cell. The motional magnetic field has a definite direction and magnitude for a time interval $\tau_c$, which is the mean time between velocity changes due to collisions of a gas particle with another particle or the wall. The parameter $\tau_c$ depends on the density, the temperature, and the collision-cross section of the gas in the measurement cell.  For a spin-1/2 system, the net effect of the randomly fluctuating field can be quantitatively calculated using a density matrix formalism~\cite{Lamoreaux3,Lamoreaux4}. For $\omega_L \tau_c\ll 1$, which is the case for the $^3$He/$^{129}$Xe co-magnetometer, the resulting frequency shift is
\begin{eqnarray}
\delta\omega_\text{m2}&=&\frac{(2\pi)^3}{9}\frac{v^2E_z^2}{c^4}\gamma^3 B_0\tau_c^2~~.
\end{eqnarray}
A false EDM effect would arise if the magnitude of the electric field would not be exactly the same after a polarity reversal. With $\Delta E_z=|E_{z,\text{up}}|-|E_{z,\text{down}}|$ (and $\Delta E_z \ll E_z$), the HV-correlated frequency shift is:
\begin{eqnarray}
\nonumber
\delta\omega_\text{m2}&=&\delta\omega_\text{m2,up}-\delta\omega_\text{m2,down}\\
&=&2\frac{(2\pi)^3}{9}\frac{v^2}{c^4}\gamma^3 B_0\tau_c^2 E_z\Delta E_z~~.
\end{eqnarray}
The values for the correlation time in the gas mixture of 30~mbar of He and 100~mbar of Xe are $\tau_{c,\text{He}}=0.37~$ns and $\tau_{c,\text{Xe}}=0.60~$ns~\cite{Lamoreaux3,Lamoreaux4}. The RMS speed values are $v_\text{He}=1575$~m/s and $v_\text{Xe}=241$~m/s. Only helium contributes significantly to this effect due to the higher RMS speed. Therefore, the corresponding false EDM by use of Eqs.~(\ref{eqn:elevel}) and (\ref{eqn:frequencydiff}) is
\begin{eqnarray}
\label{eqn:dm2}
\nonumber
d_{m2}&=&\frac{\hbar}{2E_z}\frac{\gamma_\text{Xe}}{\gamma_\text{He}}\delta\omega_\text{m2}\\
&=&\hbar\frac{(2\pi)^3}{9}\frac{v_\text{He}^2}{c^4}\gamma_\text{Xe}\gamma_\text{He}^2 B_0\tau_{c,\text{He}}^2 \Delta E_z~~.
\end{eqnarray}
Assuming (very conservatively) that the magnitudes of the electric field settings differ by 10\%, \textit{i.e.} $\Delta E_z=80$~V/cm, the quadratic term of the motional magnetic field results in a false EDM signal of:
\begin{eqnarray}
d_{m2}&=&7.6\cdot 10^{-37}~e\text{cm~~.}
\end{eqnarray}
Here, the $^3$He/$^{129}$Xe co-magnetometer benefits from a short correlation time due to the relatively high pressure.\\
\subsection{Geometric phase effect}
One of the most subtle systematic effects in any EDM experiment in which the particles are macroscopically at rest, is the influence of the geometric phase (also known as Berry's phase). The effect was originally discovered and analyzed in the context of an EDM experiment with an atomic beam of neutral atoms \cite{Commins}, and was treated extensively in the context of ultracold-neutron-based EDM experiments \cite{Lamoreaux2,Pendlebury}. The motion of particles in the plane orthogonal to the applied fields $\vect{E}$ and $\vect{B_0}$ creates a motional magnetic field according to Eq.~(\ref{eqn:motionalmagfield0}). If there is a non-zero gradient in the direction of $\vect{B_0}$, then the condition $\nabla \vect{B} = 0$ implies there must be some corresponding gradient in the radial direction with $B_\text{trans}\neq 0$. A geometric phase is caused by the collaborative action of these two types of $B_{x,y}$ components.\\
This effect is most severe in storage experiments with very low pressure, \textit{i.e.} neutron EDM experiments. The relatively high pressure in this experiment suppresses this effect substantially. In the first case (low pressure), there are no collisions of gas particles with each other.  If specular reflections at the walls allow the particles to trace out a semi-circular `orbit' around the vessel, then the combination of motional and transverse gradient fields can create an additional magnetic field shift that is linear in $E$ and differs for particles circling the vessel in opposite directions. The shift of the Larmor frequency in that case is (derived from Eqs.~(37) and (38) in \cite{Pendlebury}):
\begin{eqnarray}
\delta\omega_\text{geom,1}&=&\frac{1}{4}E_z\frac{\partial B_z}{\partial z}\frac{\gamma^2R^2}{c^2}\left(1-\frac{\omega_L^2 R^2}{0.65\cdot \frac{2}{3}v^2}\right)^{-1}~~,
\label{eqn:geom1}
\end{eqnarray}
with the RMS speed of the particles $v$ (assuming isotropic velocity distribution). Equation~(\ref{eqn:geom1}) was derived for a cylindrical cell with radius $R$, and gives an upper limit for a spherical cell with radius $R$. At a finite pressure, this effect is suppressed by a factor (\textit{cf.}~caption of Fig.~10 in \cite{Pendlebury}):
 \begin{eqnarray}
G&=&1+\left(\frac{4R^2\omega_L}{2\pi \sqrt{2/3}v\lambda}\right)^2~~.
\label{eqn:geom2}
\end{eqnarray}
Here, $\lambda$ is the mean free path of the particle in the gas mixture. In our case, this effect is dominated by geometric phases of helium due to the larger $v_\text{RMS}$ and $\lambda$. The corresponding false EDM due to geometric phases is 
\begin{eqnarray}
d_\text{geom}&=&\frac{\delta\omega_\text{geom,1,He}}{G}\frac{\gamma_\text{Xe}}{\gamma_\text{He}}\frac{\hbar}{2E_z}~~.
\label{eqn:geom3}
\end{eqnarray}
In a gas mixture of 30~mbar of He and 100~mbar of Xe, the mean free path is $\lambda_\text{He}=0.58~\mu$m, and the RMS speed values is $v_\text{He}=1575$~m/s~\cite{Chapman,Bello}. With these values,
\begin{eqnarray}
d_\text{geom}&=&1.7\cdot 10^{-31}~e\text{cm}
\end{eqnarray}
is an upper bound on the false EDM due to geometric phases.
\begin{table}
\begin{tabular}{llr}
\hline
\hline
Effect&&value / \textit{e}cm\\
\hline
Gravitational shift&$d_\text{grav}$&$8.5\cdot 10^{-33}$\\
Relax. rate shift~~~~&$d_\text{T2}$&$8.5\cdot 10^{-30}$\\
\multicolumn{3}{l}{Motional magn. field}\\
-Linear&$d_m$&$5.8\cdot 10^{-31}$\\
-Quadratic&$d_{m2}$&$7.6\cdot 10^{-37}$\\
-Geometric&$d_\text{geom}$&$1.7\cdot 10^{-31}$\\
\hline
Total (quadrature sum)&&$8.5\cdot 10^{-30}$\\
\end{tabular}
\caption{Summary of systematic false EDM effects and their estimated values. The dominant effect $d_\text{T2}$ was estimated very conservatively. \label{tab:sys}}
\end{table}
\subsection{Summary of systematic effects}
In Tab.~\ref{tab:sys}, the relevant systematic effects are summarized. The dominant contribution to the resulting total systematic error stems from $E$-field-correlated shifts in the transverse relaxation times $T_2^*$. The quadrature sum of the systematic errors is $\Delta d_\text{sys}=8.5\cdot 10^{-30}$~\textit{e}cm, a factor of 80 smaller than our current statistical uncertainty, and, therefore, does not contribute to the total uncertainty. For future experiments, the dominant contribution $d_\text{T2}$ can be further improved by more realistic (less conservative) models (\textit{e.g.}~concerning the assumed path of leakage currents).
\section{Interpretation of results}
The result for the EDM of the neutral $^{129}$Xe atom,
\begin{eqnarray}
 d_\text{Xe}&=&\left(-4.7\pm6.4\right)\cdot 10^{-28}~e\text{cm}
\end{eqnarray}
can be interpreted as an upper limit:
\begin{eqnarray}
 |d_\text{Xe}|&<&1.5 \cdot 10^{-27}~e\text{cm~(95\% C.L.)}.
\end{eqnarray}
\subsection{Limits on CP-violating observables}
In this section, we discuss the implications of the $^{129}$Xe EDM limit for possible new sources of CP violation. In establishing bounds, we make the assumption that only the source under consideration contributes to $d_\text{Xe}$. We organize the discussion by the four mechanisms that can generate an atomic EDM. These mechanisms are (i) an electron EDM, (ii) a CP-violating electron-nucleon interaction, (iii) an EDM of a valence nucleon, or (iv) a CP-violating nucleon-nucleon interaction.
\subsubsection{Limit on the electron EDM}
Measurements of the electron EDM use heavy, paramagnetic atoms or molecules which effectively enhance the interaction of $d_\text{e}$ with the applied electric field~\cite{Sandars}. Recent advances on using the exceptionally high internal effective electric field of polar molecules and ions (ThO, HfF$^+$) led to improved upper limits on the electron EDM~\cite{Cairncross,Andreev}. There is some sensitivity of diamagnetic systems to the electron EDM, although this sensitivity is very weak. The dominant contribution appears in third-order perturbation theory due to consideration of the hyperfine interaction. For the sake of completeness, we may obtain an estimate from \cite{Flambaum,Ginges} using the relation:
\begin{eqnarray}
d_\text{Xe}&=&-8 \cdot 10^{-4}d_e~~.
\end{eqnarray}
The result is
\begin{eqnarray}
|d_\text{e}|&<&1.9 \cdot 10^{-24}~e\text{cm~(95\% C.L.)}~~.
\end{eqnarray}
\subsubsection{Limits on CP-violating electron-nucleon interactions}
CP-violating electron-nucleon interactions can be classified as scalar-pseudoscalar, pseudoscalar-scalar and tensor interactions with dimensionless coupling constants  $C_N^{SP}$, $C_N^{PS}$ and $C_N^{T}$ for the nucleon $N$, respectively. Their contributions to the atomic EDM according to \cite{Ginges,Dzuba1} are
\begin{eqnarray}
\nonumber
d_\text{Xe}&=&(-5.6\cdot10^{-23}C_N^{SP}+1.6 \cdot 10^{-23}C_N^{PS}\\
&&+5.7 \cdot 10^{-21}C_N^T)\langle \sigma_N\rangle~e\text{cm}~~,
\end{eqnarray}
where $\langle \sigma_N\rangle$ is the neutron ($N=n$) or proton ($N=p$) polarization in the $^{129}$Xe nucleus, which can be determined from shell-model calculations~\cite{Dzuba1}: 
the magnetic moment $\mu_\text{Xe}$ of the $^{129}$Xe nucleus is composed entirely from the spin magnetic moment of the valence neutron and the spin magnetism of the polarized nuclear core, giving $\mu_\text{Xe} = \mu_n \langle\sigma_n\rangle+\mu_p\langle \sigma_p\rangle$ with 
\begin{eqnarray}
\nonumber \langle\sigma_\text{n}\rangle&=&0.76\\
\langle\sigma_\text{p}\rangle&=&0.24
\end{eqnarray}
and $\mu_n$ and $\mu_p$ being the magnetic moments of the neutron and the proton. For the $^{129}$Xe nucleus, $\langle\sigma_\text{n}\rangle+\langle\sigma_\text{p}\rangle=1$ holds. These numbers are used to extract limits on the CP-odd electron-nucleon interaction originating from the neutron and proton. The results are summarized in Tab.~\ref{tab:CPlimits}.\\
\subsubsection{Limits on CP-violating nucleon-nucleon interactions and intrinsic nucleon EDMs}
The $^{129}$Xe atom is sensitive to all the CP-violating nuclear observables through its nuclear Schiff moment $S$, which measures the detectable, unshielded part of a nuclear EDM \cite{Schiff,Liu,Senkov}. The relationship between the nuclear Schiff moment and the atomic EDM is fairly well understood. Recent results \cite{Dzuba2,Dzuba1,Yoshinaga} of atomic structure calculations give:
\begin{eqnarray}
d_\text{Xe}&=& 3.8 \cdot 10^{-18}~\frac{S}{e~\text{fm}^3}~e\text{cm}~~.
\end{eqnarray}
From our measurement result, we derive the following upper limit:
\begin{eqnarray}
|S_\text{Xe}|&<& 4.0 \cdot 10^{-10}~e~\text{fm}^3\text{~~(95\% C.L.)}~~.
\label{eqn:Schiff}
\end{eqnarray}
The different contributions to the Schiff moment are: intrinsic nucleon EDMs and CP-violating nucleon-nucleon interactions.\\
The intrinsic neutron EDM $d_\text{n}$ and proton EDM $d_\text{p}$ give rise to a measurable Schiff moment of \cite{Dzuba0}:
\begin{eqnarray}
S_\text{Xe}&=&0.63~ \text{fm}^2 d_\text{n} + 0.125~ \text{fm}^2 d_\text{p}~~.
\end{eqnarray}
This relation can be used to extract the upper bounds on the neutron and proton EDM from Eq.~(\ref{eqn:Schiff}),
\begin{eqnarray}
 |d_\text{n}|&<&6.4 \cdot 10^{-23} ~e\text{cm~~(95\% C.L.)}
\end{eqnarray}
and
\begin{eqnarray}
|d_\text{p}| &<&3.2 \cdot 10^{-22} ~e\text{cm~~(95\% C.L.)}~~.
\end{eqnarray}
The largest contribution to the atomic Xe EDM is expected to arise through CP-violating nucleon-nucleon interaction. The exchange of a $\pi_0$-meson is the most efficient mechanism of generating CP-violating nuclear forces (due to the large coupling constant, the small pion mass, and large differences in the outer proton and neutron orbitals in heavy nuclei). These couplings are classified by their isotopic properties, \textit{i.e.} isoscalar, isovector and isotensor coupling with constants $g_0$, $g_1$, and $g_2$, respectively.
A calculation of the Schiff moment, including a full account of core polarization effects that were found to have a large effect (see Tab.~V in \cite{Dmitriev}), yields:
\begin{eqnarray}
S&=&(0.008 g_0 + 0.006 g_1 - 0.009 g_2)~e~\text{fm}^3~~.
\end{eqnarray}
It should be noted that there is considerable disagreement between various calculations of $S_\text{Xe}(g_0,g_1,g_2)$. To set limits on $g_{0,1,2}$, we used the quoted best values for $^{129}$Xe from recent reviews~\cite{Engel,Chupp}. The corresponding upper limits are
\begin{eqnarray}
\nonumber
|g_0|&<&5.0\cdot10^{-8}\\
\nonumber
|g_1|&<&6.7\cdot10^{-8}\\
|g_2|&<&4.5\cdot10^{-8}~\text{(95\% C.L.)}~~.
\end{eqnarray}
\begin{table}
\begin{tabular}{l|l|lr|l}
\hline
\hline
Param.&Limit (this work)&\multicolumn{2}{|l|}{Best limit (other work)}&Theory\\
\hline
$d_\text{Xe}$&$1.5\cdot10^{-27}~e$cm&$4.8\cdot10^{-27}~e$cm&\cite{Sachdeva}&\\
\hline
$d_e$&$1.9\cdot10^{-24}~e$cm&$1.1\cdot10^{-29}~e$cm&\cite{Andreev}&\cite{Flambaum,Ginges}\\
$C_\text{n}^{SP}$&$3.6\cdot10^{-5}$&$1.3\cdot10^{-8}$&\cite{Graner2}&\cite{Ginges}\\
$C_\text{n}^{PS}$&$1.3\cdot10^{-4}$&$1.2\cdot10^{-7}$&\cite{Graner2}&\cite{Dzuba1}\\
$C_\text{n}^{T}$&$3.5\cdot10^{-7}$&$1.5\cdot10^{-10}$&\cite{Graner2}&\cite{Dzuba1}\\
$C_\text{p}^{SP}$&$1.1\cdot10^{-4}$&-&&\cite{Ginges}\\
$C_\text{p}^{PS}$&$4.0\cdot10^{-4}$&-&&\cite{Dzuba1}\\
$C_\text{p}^{T}$&$1.1\cdot10^{-6}$&-&&\cite{Dzuba1}\\
$S_\text{Xe}$&$4.0\cdot10^{-10}~e$~fm$^3$&$1.3\cdot10^{-9}~e$~fm$^3$&\cite{Sachdeva}&\cite{Dzuba2,Dzuba1}\\
$d_n$&$6.4\cdot10^{-23}~e$cm&$1.6\cdot10^{-26}~e$cm&\cite{Graner2}&\cite{Dzuba0}\\
$d_p$&$3.2\cdot10^{-22}~e$cm&$2.0\cdot10^{-25}~e$cm&\cite{Graner2}&\cite{Dzuba0}\\
$g_0$&$5.0\cdot10^{-8}$&$2.3\cdot10^{-12}$&\cite{Graner2}&\cite{Dmitriev,Yoshinaga}\\
$g_1$&$6.7\cdot10^{-8}$&$1.1\cdot10^{-12}$&\cite{Graner2}&\cite{Dmitriev,Yoshinaga}\\
$g_2$&$4.5\cdot10^{-8}$&$1.1\cdot10^{-12}$&\cite{Graner2}&\cite{Dmitriev,Yoshinaga}\\
\hline
\hline
\end{tabular}
\caption{Summary of limits on different sources of CP violation, extracted from the $^{129}$Xe EDM limit of this work. Each limit is based on the assumption that it is the sole contribution to the atomic EDM. All limits are 95\% confidence limits (theoretical uncertainties are not included). Further, we added best limits from other work (mostly derived from the $^{199}$Hg-EDM measurement in \cite{Graner2}). \label{tab:CPlimits}}
\end{table}
\subsection{Axion limits}
The exchange of an axion-like particle between atomic electrons (e) and the nucleus (N) may induce EDMs of atoms and molecules. This interaction is described by a CP violating potential (Yukawa-type) which depends on the product of a scalar $g_e^s$ and a pseudoscalar $g_N^p$ coupling constant. The contribution to the EDM of $^{129}$Xe was calculated in~\cite{Dzuba3} where the interaction with the specific combination of these constants, $g_e^s g_N^p$ was considered, \textit{i.e.}, the interaction of the non-zero nuclear spin with the closed electron shell of xenon. For axion masses $m_a<1$~keV, the interaction becomes long-range, \textit{i.e.}, $\lambda_c= h/(m_a c) \gg r_\text{Xe}$ (atomic radius), (see Eq.~(8) in~\cite{Dzuba3}), and the induced atomic EDM becomes independent of $m_a$. The asymptotic value for the Xe EDM is
\begin{eqnarray}
d_\text{Xe} &=& 1.5\cdot10^{-13}~e\text{cm}~|g_e^s g_N^p|~~,
\end{eqnarray}
and we derive the following upper limit:
\begin{eqnarray}
|g^s_e g^p_N| < 1\cdot 10^{-14}~\text{for}~m_a<1~\text{keV}~.
\end{eqnarray}

\section{Conclusion and Outlook}
We improved the limit on the permanent EDM of the $^{129}$Xe atom by a factor of 3 using the detection of free spin precession of co-located gaseous, nuclear polarized $^3$He and $^{129}$Xe samples with a SQUID as magnetic flux detector. $^3$He is used as co-magnetometer to render the experiment insensitive to drifts and fluctuations of the magnetic guiding field ($\approx 400$~nT) inside a magnetically shielded room. The experiment's EDM sensitivity strongly benefits from the long spin-coherence times of several hours reached in $^3$He/$^{129}$Xe gas mixtures at total pressures around 100~mbar. From our experimental result $d_\text{Xe}=\left(-4.7\pm6.4\right)\cdot 10^{-28}~e\text{cm}$, we place a new upper limit on the $^{129}$Xe EDM of $|d_\text{Xe}|<1.5 \cdot 10^{-27}$~\textit{e}cm~(95\% C.L.).\\
The EDM sensitivity of our experiment can be significantly improved with the next optimization steps:\\
the use of external electrodes forced us to apply a modest electric field of 800~V/cm in order to ensure that the same electric field strength could be maintained within the insulating spherical EDM glass cell over the duration of a single measurement run. A modified EDM cell, still spherical, but with integrated silicon electrodes will allow us to increase the electric field by a factor of 3 to 5, and with it the measurement sensitivity, accordingly.\\
Efforts to improve the magnetic field homogeneity and the shielding factor of the MSR are essential. This will not only have a positive effect on the duration of our spin coherence times $T_2^*$, which are currently limited by field gradients. A new, three-layer mu-metal shielded room with a better overall shielding factor than the previous one at the research center J\"{u}lich is currently under development. This allows the inner mu-metal cylinder near the EDM spectrometer to be removed, which currently worsens our system noise by a factor of 10, and thus, also the signal-to-noise ratio.\\
With these measures, the currently achievable statistical Xe-EDM sensitivity of  $6\cdot10^{-28}$~\textit{e}cm per day (see Tab.~\ref{tab:result}) can be improved down to values that are similar to the one from the most recent $^{199}$Hg-EDM experiment with $4\cdot10^{-29}$~\textit{e}cm per day~\cite{Graner1}. The present upper EDM limit on the $^{199}$Hg atom, $|d_\text{Hg}|<7.4 \cdot 10^{-30}$~\textit{e}cm (95\% C.L.), to date provides the tightest constrains on the CP-violating observables in atoms, and the derived limit on $d_n$ surpasses the current best limit measured with free neutrons~\cite{Baker}. Here, the diamagnetic $^{129}$Xe atom provides a complementary system more sensitive to proton parameters which is needed to complete the picture of CP violation.
\appendix
\section*{Acknowledgments}
We owe special thanks to O.~Grasdijk (Van Swinderen Institute, University of Groningen, The Netherlands; now at Yale University) for his help and valuable contributions during our measurement campaigns; and also his coil design and construction. 
We further thank the following persons: L.~Willmann and K.~Jungmann for many helpful and inspiring discussions (both Van Swinderen Institute, University of Groningen, The Netherlands); R.~Jera (glass blower) and P.~Bl\"{u}mler (NMR-detection) (both Uni Mainz) and V.~Angelov (pA-meters) (Uni Heidelberg). This project was funded by the Deutsche Forschungsgemeinschaft (DFG) under grants HE~2308/12-1 as well as SCHM~2708/3-1, and Carl-Zeiss-Stiftung. We appreciate the generous financial support of the Research Center J\"{u}lich and the provision of the infrastructure there (magnetic shielded room and laboratories) to conduct the experiment. The Mainz cluster of excellence PRISMA "Precision Physics, Fundamental Interactions and Structure of Matter" is also greatly acknowledged for bridge financing this project.

\section{Electric field measurement}
\label{sec:efieldmeasurement}
\begin{figure}
\includegraphics[width=\columnwidth]{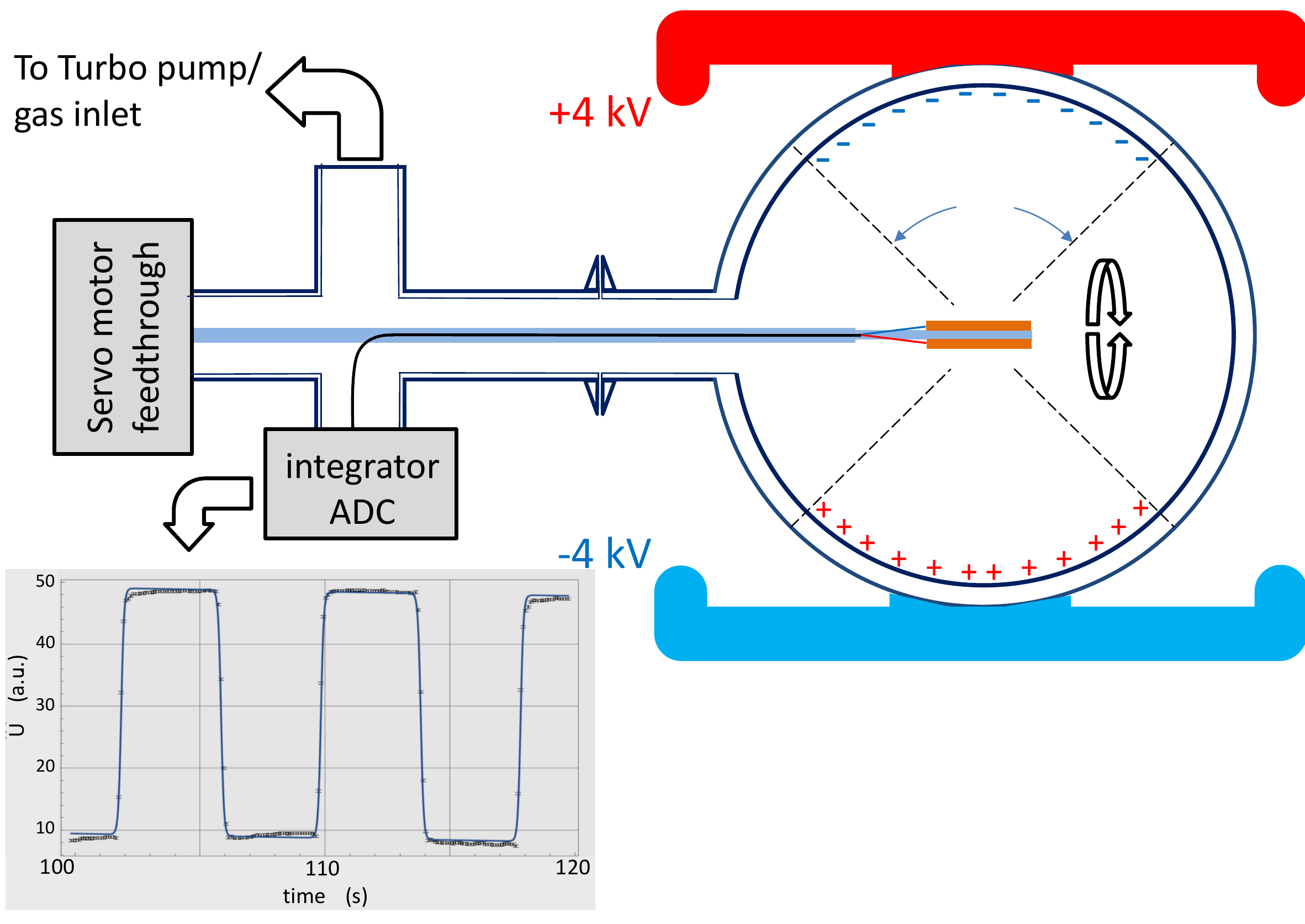}
\caption{Sketch of the field mill assembly to measure the effective electric field inside a spherical glass vessel in between two electrodes kept at +~4~kV and -~4~kV, respectively. Two copper plates at its center which form a plate capacitor are rotated by 180$^\circ$ back and forth every 4~s. The displacement currents are monitored by means of an integrator circuit as shown on the bottom left. Opposing electric fields can build up due to charge separation on the inner surfaces of the insulator (glass), which leads to a change in capacitance (see text). 
\label{fig:fieldmill}
}
\end{figure}
\begin{figure}
\includegraphics[width=\columnwidth]{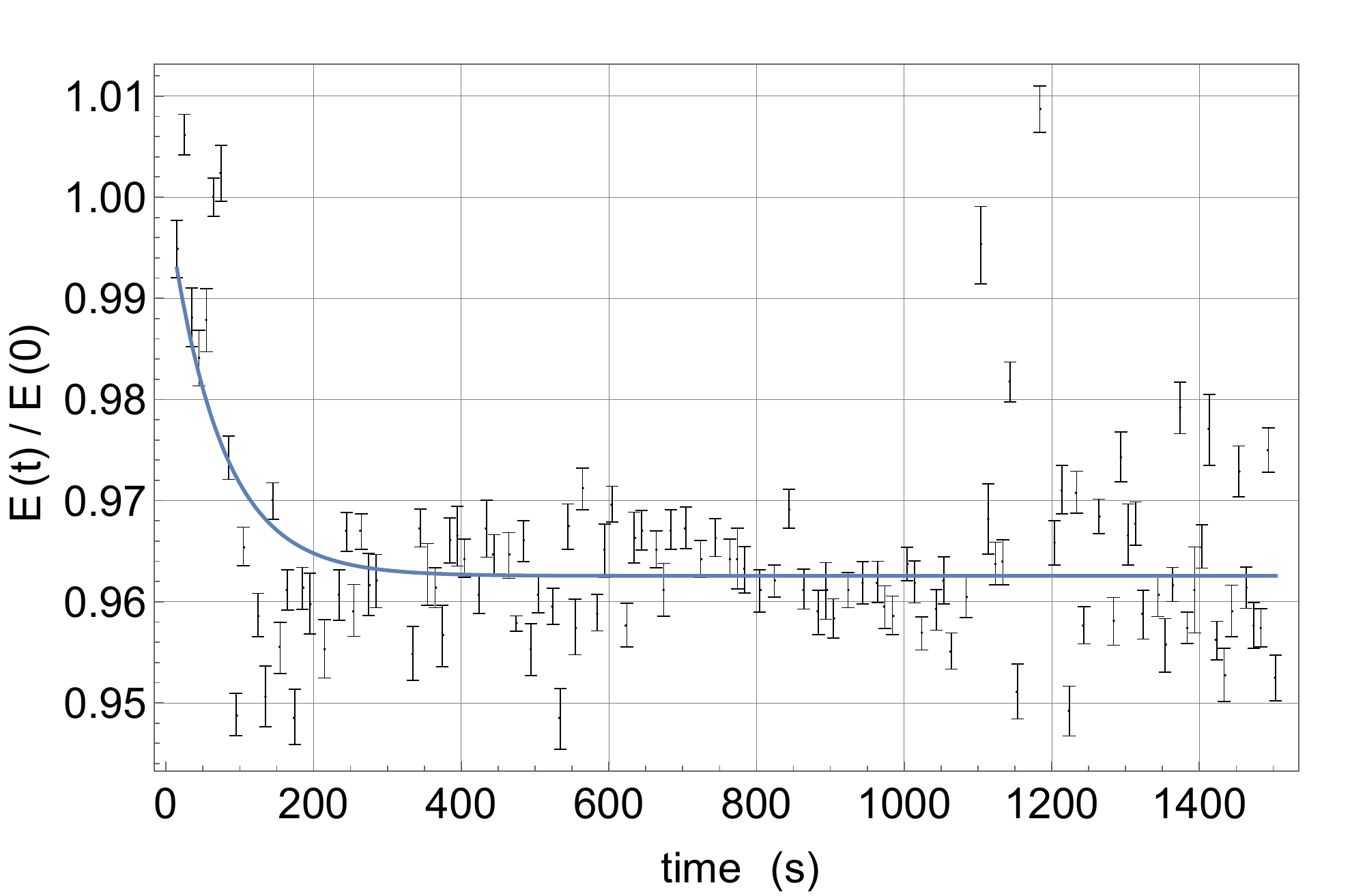}
\caption{Time course of the electric field (normalized to $E(t=0)$) inside the spherical glass vessel as recorded by the field mill. The cell was filled with a $^3$He/$^{129}$Xe gas mixture of 30~mbar/100~mbar. After an initial field drop, the field stabilizes at $\approx 96\%$ of the field at $t=0$, the value that was measured without glass cell. The externally applied field was 800~V/cm. Each data point with error bar results from an amplitude fit to five rectangular integrator pulses in succession (duration: 20~s). The temporal course of the data points show a somewhat higher fluctuation presumably due to environmental electrostatic disturbances. The outliers at about 1200 s are artifacts caused by resets of the integrator circuit (which are needed from time to time in order to keep the integrator output voltage within the dynamic range).
\label{fig:Efieldvstime}
}
\end{figure}
The use of spherically shaped glass vessels for the sample spins immersed in the homogeneous electric field between the two electrodes (plate capacitor) demands the control of the electric field inside the cell. As charges can accumulate at different locations on the inner and outer surfaces of a glass cell, the electric field seen by the $^3$He and $^{129}$Xe atoms may decrease over time and eventually vanish in case an opposing electric field builds up, which compensates the outer field. To quantify the (time-dependent) effective electric field inside the EDM cell, two different field sensors were developed for off-line measurements. The first method based on a birefringent lithium niobate electro-optic crystal with optical fiber read out is discussed in detail in~\cite{Grasdijk}. Here we present the results obtained with the second setup shown in Fig.~\ref{fig:fieldmill}, the so-called field mill: the spherical glass cell is put in between the electrodes which were also used in the EDM runs. Care was taken that the poles of the glass cell were in good electrical contact with the electrodes. As with the EDM run, conductive foam was used which, slightly pressed, molds to both surfaces (contact area $\approx 4~\text{cm}^2$). Prior to that, the inner and outer surfaces have been thoroughly cleaned following the cleaning procedure of the EDM cells~\cite{Repetto}: a 1:5 mixture of Mucasol (Sch\"{u}lke und Mayr GmbH, Norderstedt, Germany) and distilled water. After that the cell was dried in a vacuum oven at 80~$^\circ$C for at least 12~h. The spherical glass cell as shown in Fig.~\ref{fig:fieldmill} is further connected to a crosspiece which in turn is connected to a Turbo-pump station, the gas in- and outlet, as well as a vacuum feed-through for the signal supply lines and a mechanical rotary feed-through. Via the vacuum rotary feed-through a servo-motor rotates a thin glass tube (tapered towards the centre of the cell) 180$^\circ$ back and forth. At its very end, two copper plates ($5\times 10$~mm$^2$) are fixed which form a plate capacitor to measure the displacement currents proportional to the effective electric field inside the glass cell. The signal lines are fed outside to an integrator circuit with ADC board.
Figure~\ref{fig:fieldmill} shows the output of the integrator circuit at 180$^\circ$ rotations after every 4 seconds. The applied electric field was $E_z=800$~V/cm. The charge data was recorded both with the glass cell attached and with the cell removed. With the cell connected and after beeing pumped for several days, a $^3$He/$^{129}$Xe gas mixture of 30~mbar/100~mbar was filled in (typical EDM run conditions, see Tab.~\ref{tab:result}). Figure~\ref{fig:Efieldvstime} shows the time dependence of the effective electric field inside the cell normalized to the measured field amplitude at $t=0$ (time at which the HV was applied to the electrodes). After a short field relaxation (the signal drop shows an exponential behavior) the field inside the cell stabilizes above 95\% of the initial field (fit-curve to the data points). In total, the temporal behavior of the field was recorded for 16 hours. As time-averaged field  value inside the EDM cell we measured  $(96.6\pm0.1)$\% of the initial field value at an externally applied electric field of $E_z=800$~V/cm. Within the error bars, the initial field amplitude agrees with the one measured without glass cell.\\
\begin{figure}
\includegraphics[width=\columnwidth]{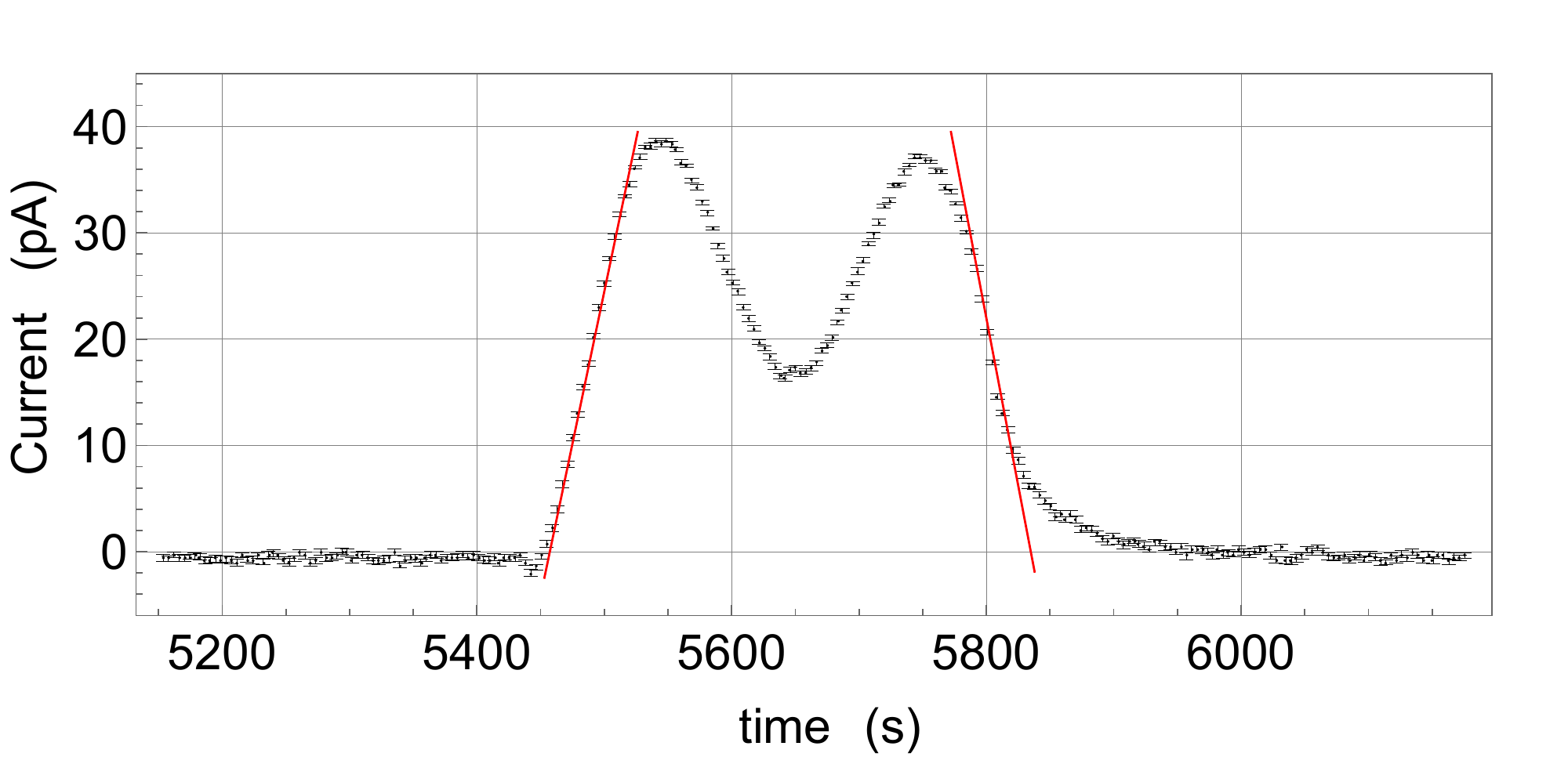}
\caption{Displacement current measured during an electric field reversal (-4~kV to +~4 kV) by means of the pA-meter in the HV supply lines (see Fig.~\ref{fig:cell}). The voltage ramping rate is 25~V/s with an interruption of 80~s at $V=0$~kV to switch the relays, \textit{i.e.}, a total of 400~s for polarity reversal. A built-in low pass filter causes the double hump structure (the slope of the wings should be the same indicated by the red lines). The exponential relaxation of the displacement current after $t=5800$~s  is due to an electric field drop of $\approx3\%$ which is associated with a capacitance change (details see text). 
\label{fig:pAdata2}
}
\end{figure}
This temporal behavior of the electric field could also be observed on-line during the EDM runs by means of the pA-meters installed in the HV-supply lines (see Fig.~\ref{fig:cell}). If there is a measurable decrease in the electric field inside the cell volume, an opposing electric field must build up through charge separation (\textit{e.g.} Townsend-type gas discharges), as indicated in Fig.~\ref{fig:fieldmill}. This causes a gradual increase of the total capacitance of the electrode system, leading to enlarged displacement currents which can be monitored by the pA-meters. The capacitance of the EDM-electrode assembly (including the cell) is about 1.5 pF (see Fig.~\ref{fig:pAdata2}). If there is charge separation which completely compensates the field inside the cell, the capacitance rises to approximately 12 pF. This was calculated with Comsol-Multiphysics using the exact geometry of the EDM assembly, the relative permittivity of aluminosilicate glass ($\epsilon_r\approx6$) and wall thickness $d=2.5$~mm of the spherical cell. For $\alpha/2>30^\circ$ which characterizes the distribution of charges around the inner pole caps of the cell (see Fig.~\ref{fig:fieldmill}), the capacitance reaches a plateau at around 12 pF. Assuming an exponential relaxation of the field inside the cell, the temporal behavior of the total capacitance $C$ can be written as:
\begin{eqnarray}
C(t)&=&1.5~\text{pF}\cdot\left(1-X(t)\right)+ 12~\text{pF}\cdot X(t)~~,
\end{eqnarray}
with $X(t)=X_0(1-\exp(-t/\tau))$, where $X_0$ denotes the decreasing proportion of the field and $\tau$ is the characteristic time constant. The displacement current then gives:
\begin{eqnarray}
i(t)&=&U\frac{\text{d} C}{\text{d} t}=U\frac{X_0}{\tau}\cdot 10.5~\text{pF}\cdot \exp(-t/\tau)
\end{eqnarray}
For $X_0=0.03$, $\tau=200$~s and $U=8$~kV as was encountered off-line (see Fig.~\ref{fig:Efieldvstime}) we therefore expect a displacement current of $i(t)\approx 13~\text{pA}\cdot \exp(-t/200~\text{s})$ which should be clearly visible (on-line) after each electric-field reversal. Figure~\ref{fig:pAdata2} shows a detail (polarity reversal) from Fig.~\ref{fig:pAdata} around $t\approx5600$~s: electric-field reversal monitored by the pA-meters manifests in a double hump. The tail of the hump is smeared out which can be attributed to the exponential relaxation of the electric field. Here, one reads $\approx10$~pA for the current amplitude and $\tau\approx150$~s in excellent agreement with the off-line results. Therefore, relative electric field drops larger than 5\% can definitely be excluded for all EDM runs. 
\cleardoublepage
\section{Orthogonalization of fit function}
\label{sec:ortho}
The appropriate function that includes all deterministic phase shifts (chemical shift and Earth's rotation by a linear term, as well as the Ramsey-Bloch-Siegert shift described by four exponential terms) and contains the parametrization of an EDM induced phase shift, is given by Eq.~(\ref{eqn:edmfit}). Fitting this function to the weighted phase difference data causes numerical problems inside the fitting routine due to a very high correlation of fit parameters.  For example, in run number 6 (see Tab.~\ref{tab:result}) the correlation matrix is:
\vspace{5mm}\\
{\footnotesize $
\begin{array}{c|cccccc}
 &\Phi_0&\Delta\omega_\text{lin}&E_\text{He}&E_\text{Xe}&F_\text{Xe}&F_\text{He}\\
\hline
\Phi_0 & 1. & -0.9999 & -0.9999 & -0.9969 & 0.9998 & 0.9877 \\
\Delta\omega_\text{lin} & -0.9999 & 1. & 0.9999 & 0.9964 & -0.9996 & -0.9867 \\
E_\text{He} & -0.9999 & 0.9999 & 1. & 0.9974 & -0.9999 & -0.9887 \\
E_\text{Xe} & -0.9969 & 0.9964 & 0.9974 & 1. & -0.9983 & -0.9968 \\
F_\text{Xe}& 0.9998 & -0.9996 & -0.9999 & -0.9983 & 1. & 0.9906 \\
F_\text{He} & 0.9877 & -0.9867 & -0.9887 & -0.9968 & 0.9906 & 1. \\
\end{array}
$}\\
\vspace{5mm}\\
All entries are very close to $\pm 1$. Inside the fitting routine a lot of matrix inversions of almost singular matrices have to be calculated which causes numerical errors and instabilities. And, as a consequence, the optimum is not found reliably.\\
The solution is to rewrite the fit function, so that the individual terms are orthogonal to each other.  The fit function has to be the sum of orthogonal terms $f_i(t)$ (multiplied by the fit parameters). In this case, orthogonal is defined as
\begin{eqnarray}
\label{eqn:ortho}
\int_{t_B}^{t_E} \! \exp(-2 \cdot t / T_{2, \text{Xe}}^*) f_i(t) \cdot f_j(t) \, \mathrm{d}t &\propto&\delta_{i,j}~.
\end{eqnarray}
Here, $t_B$ is the start time, and $t_E$ is the stop time of the measurement run. By defining the inner product with a weighing function, one takes into account that data points in the beginning have a higher weight than the ones at the end, due to the increasing phase error (decreasing xenon amplitude). In order to convert the terms of Eq.~(\ref{eqn:edmfit}) to orthogonal terms, one can  use the Gram-Schmidt process~\cite{numrec}. This has to be done for every run anew, as $T_{2, \text{Xe}}^*$, $T_{2, \text{He}}^*$ and the length of the run varies.\\
For the given example, the result is (numerical values rounded):
\begin{eqnarray} 
\label{eqn:fitmodel2} 
\Delta\Phi(t)&=&a_0\nonumber\\
&&+ a_1 \cdot \left(t-4927.72\right)\nonumber\\
&& + a_2 \cdot \left(\exp(-t/T_{2, He}^*)+ 0.000013 \cdot t - 0.99 \right)\nonumber\\
&& + a_3 \cdot \left(\exp(-t/T_{2, Xe}^*)-21.57 \cdot \exp(-t/T_{2, He}^*) -0.00023 \cdot t +  20.58\right)\nonumber \\
&& + a_4 \cdot \left(\exp(- 2 \cdot t/T_{2, He}^*)-0.031 \cdot \exp(-t/T_{2, Xe}^*)-2.85 \cdot \exp(-t/T_{2, He}^*) - 0.000015 \cdot t + 1.88\right)\nonumber\\
&& + a_5 \cdot \left(\exp(- 2 \cdot t/T_{2, Xe}^*)+172.64 \cdot \exp(- 2 \cdot t/T_{2, He}^*)-9.42 \cdot \exp(-t/T_{2, Xe}^*) -446.63 \cdot  \exp(-t/T_{2, He}^*)\right)\nonumber\\
&&~
\end{eqnarray}
with a corresponding correlation matrix:
\vspace{5mm}\\
{\footnotesize $
\begin{array}{c|cccccc}
  &a_0&a_1&a_2&a_3&a_4&a_5\\
\hline
a_0 & 1. & -0.12 & 0.09 & 0.05 & -0.03 & 0.02 \\
a_1 & -0.12 & 1. & -0.11 & -0.08 & 0.04 & -0.03 \\
a_2 & 0.09 & -0.11 & 1. & 0.12 & -0.07 & 0.05 \\
a_3 & 0.05 & -0.08 & 0.12 & 1. & -0.11 & 0.09 \\
a_4 & -0.03 & 0.04 & -0.07 & -0.11 & 1. & -0.11 \\
a_5 & 0.02 & -0.03 & 0.05 & 0.09 & -0.11 & 1. \\
\end{array}
$}\\
\vspace{5mm}\\
With this fit function, the correlation between the fit parameters was greatly reduced and the fitting routine worked reliably. In order to  investigate the influence of experimental parameters (\textit{e.g.} $T_a$) on the correlation (cf. Fig.~\ref{fig:EDMerrorvsta}), the EDM phase term was not included in the orthogonalization process, but rather added subsequently. Unavoidably, this increased the correlation between
all the fit parameters (especially between the EDM phase term and the four exponential terms describing the Ramsey-Bloch-Siegert shift). There is no physical effect causing correlation in this case, but the correlation stems from similar time dependent signals  which are not orthogonal to each other in the sense of Eq.~(\ref{eqn:ortho}). However, the correlation is significantly less than one, posing no numerical challenge to the fitting routine. 
Including the EDM term, the correlation matrix for this example is:
\vspace{5mm}\\
{\footnotesize $
\begin{array}{c|ccccccc}
   &a_0&a_1&a_2&a_3&a_4&a_5&g\\
\hline
a_0 & 1. & -0.56 & 0.57 & 0.57 & 0.43 & -0.49 & -0.58 \\
a_1 & -0.56 & 1. & -0.83 & -0.84 & -0.66 & 0.74 & 0.87 \\
a_2 & 0.57 & -0.83 & 1. & 0.89 & 0.70 & -0.78 & -0.92 \\
a_3 & 0.57 & -0.84 & 0.89 & 1. & 0.71 & -0.79 & -0.94 \\
a_4 & 0.43 & -0.66 & 0.70 & 0.71 & 1. & -0.70 & -0.77 \\
a_5 & -0.49 & 0.74 & -0.78 & -0.79 & -0.70 & 1. & 0.85 \\
g & -0.58 & 0.87 & -0.92 & -0.94 & -0.77 & 0.85 & 1. \\
\end{array}
$}

\pagebreak



\begin{thebibliography}{100}
\bibitem{Khriplovich} I.~B.~Khriplovich, Phys. Lett. B {\bf 173}, 193 (1986).
\bibitem{Pospelov} M. Pospelov and A. Ritz, Phys. Rev. D {\bf 89}, 056006 (2014).
\bibitem{Chupp} T.~E. Chupp, P. Fierlinger, M.~J. Ramsey-Musolf, and J.~T. Singh, Rev. Mod. Phys. {\bf 91}, 015001 (2019).
\bibitem{Jungmann} K. Jungmann, Ann. Phys. (Berlin) {\bf 525}, 550 (2013).
\bibitem{Chupp2} T.~Chupp, M.~J. Ramsey-Musolf, Phys. Rev. C {\bf 91} (2015).
\bibitem{Engel} J. Engel, M.~J. Ramsey-Musolf, U. van Kolck, Progress in Particle and Nuclear Physics {\bf 71} (2013) 21.
\bibitem{Baker} C.~A. Baker \textit{et al.} (RAL/Sussex/ILL collaboration ), Phys. Rev. Lett. {\bf 97}, 131801, (2006).
\bibitem{Graner1} B. Graner, Y. Chen, E.~G. Lindahl, and B.~R. Heckel, Phys. Rev. Lett. {\bf 116}, 161601 (2016). 
\bibitem{Graner2} B. Graner, Y. Chen, E.~G. Lindahl, and B.~R. Heckel, Phys. Rev. Lett. {\bf 119}, 119901 (2017). 
\bibitem{Andreev} V. Andreev \textit{et al.} (ACME Collaboration), Nature {\bf 562}, 355 (2018).
\bibitem{Cairncross} W.~B. Cairncross, D.~N. Gresh, M. Grau, K.~C. Cossel, T.~S. Roussy, Y.~Ni, Y.~Zhou, J.~Ye, and E.~A. Cornell, Phys. Rev. Lett. {\bf 119}, 153001 (2017).
\bibitem{Sachdeva} N. Sachdeva \textit{et al.}, arXiv:1902.02864v1 (2019). 
\bibitem{Rosenberry} M.~A. Rosenberry and T.~E. Chupp, Phys. Rev. Lett {\bf 86}, 22 (2001).
\bibitem{Dzuba3} V.~A. Dzuba, V.~V. Flambaum, I.~B. Samsonov, and Y.~V. Stadnik, Phys. Rev. D {\bf98}, 035048 (2018)
\bibitem{Schiff} L.~I. Schiff, Phys. Rev. {\bf 132}, 2194 (1963).
\bibitem{FlambaumKozlov} V.~V. Flambaum and A. Kozlov, Phys. Rev. A {\bf 85}, 022505 (2012).
\bibitem{Gemmel} C. Gemmel \textit{et al.}, Eur. Phys. J. D {\bf 57}, 303 (2010). 
\bibitem{Kay} S.~M. Kay, {\it Fundamentals of Statistical Signal Processing: Estimation Theory} (Prentice Hall, New Jersey, 1993).
\bibitem{Cates} G.~D. Cates, S.~R. Schaefer, and W. Happer, Phys. Rev. A {\bf 37}, 8 (1988).
\bibitem{Paschen} F. Paschen,  Wied. Ann. {\bf 37}, 69 (1889).
\bibitem{Lee} S.-K. Lee and M.~V. Romalis, J. Appl. Phys. {\bf 103}, 084904 (2008).
\bibitem{Zimmer} S. Zimmer, Ph.D. thesis, Johannes Gutenberg University Mainz, 2018.
\bibitem{Grasdijk} O. Grasdijk, Ph.D. thesis, University of Groningen, 2018.
\bibitem{Magnicon} Magnicon GmbH, Hamburg, Germany.
\bibitem{Cryoton} CRYOTON Co. Ltd., Russia.
\bibitem{ADS1299} ADC chip (ADS1299) from Texas Instruments, a simultaneously-sampling, 24-bit, delta-sigma converter.
\bibitem{Repetto} M. Repetto, E. Babcock, P. Bl\"{u}mler, W. Heil, S. Karpuk, and K. Tullney, J. Magn. Reson. {\bf 252}, 163 (2015).
\bibitem{Schmiedeskamp} J. Schmiedeskamp, W. Heil, E.W. Otten, R.K. Kremer, A. Simon, and J. Zimmer, Eur. Phys. J. D {\bf 38}, 427 (2006).
\bibitem{Rich} D.~R. Rich, T.~R. Gentile, T.~B. Smith, A.~K. Thompson, and G.~L. Jones, Appl. Phys. Lett. {\bf 80}, 2210 (2002).
\bibitem{Iseg} ISEG Spezialelektronik GmbH, Radeberg, Germany.
\bibitem{Karpuk} S. Karpuk  \textit{et al.}, Phys. Part. Nucl. {\bf 44}, 904 (2013).
\bibitem{Wolf} M. Wolf, Ph.D. thesis, Johannes Gutenberg University Mainz, 2004.
\bibitem{Hiebel} S. Hiebel, T. Gro\ss mann, D. Kiselev, J. Schmiedeskamp, Y. Gusev, W. Heil, S. Karpuk, J. Krimmer, E. W. Otten, and Z. Salhi, J. Magn. Reson. {\bf 204}, 37 (2010).
\bibitem{Thien} F. Thien \textit{et al.}, Respirology {\bf 13}, 599 (2008).
\bibitem{Appelt} S. Appelt, A. B. A. Baranga, C. J. Erickson, M. V. Romalis, A. R. Young, and W. Happer, Phys. Rev. A {\bf 58}, 1412 (1998).
\bibitem{Thiel} F. Thiel \textit{et al.}, Rev. Sci. Instrum. {\bf 78}, 035106 (2007).
\bibitem{Altarev} I. Altarev \textit{et al.}, Rev. Sci. Instrum. {\bf 85}, 075106 (2014).
\bibitem{NelderMead} J. A. Nelder, R. Mead, Computer Journal {\bf 7}, 308 (1965).
\bibitem{Allmendinger} F. Allmendinger \textit{et al.}, Eur. Phys. J. D {\bf 71}, 98 (2017).
\bibitem{Beringer} J. Beringer \textit{et al.} (Particle Data Group), Phys. Rev. D {\bf 86}, 010001 (2012).
\bibitem{Bloch} F. Bloch and A. Siegert, Phys. Rev. {\bf 57}, 522 (1940). 
\bibitem{Ramsey} N .F. Ramsey, Phys. Rev. {\bf 100}, 1191 (1955). 
\bibitem{Flowers} J. L. Flowers, B. W. Petley, and M. G. Richards, Metrologia {\bf 30}, 75 (1993).
\bibitem{Pfeffer} M. Pfeffer, O. Lutz, J. Magn. Reson. A {\bf 108}, 106 (1994).
\bibitem{Makulski} W. Makulski, Magn. Reson. Chem. {\bf 53}, 273 (2015).
\bibitem{Jackson} J. D. Jackson, Classical electrodynamics (John Wiley and Sons, 1998).
\bibitem{Caciagli}A. Caciagli, R.~J. Baars, A.~P. Philipse, and  B.~W.~M. Kuipers, J. Magn. Materials {\bf 456}, 423 (2018).
\bibitem{Allan} D. W. Allan,  Proc. of IEEE {\bf 54}, 221, (1966).
\bibitem{Allan2} D. W. Allan, Proc. Sixth Ann. Precise Time and Time Interval Planning Meet., Washinton, DC, 109 (1974).
\bibitem{Barnes} J. A. Barnes \textit{et al.}, IEEE Trans. Instrum. Meas. {\bf 20}, 105 (1971).
\bibitem{Lesage} P. Lesage, C. Audoin, IEEE Trans. Instrum. Meas. {\bf 22}, 157 (1973).
\bibitem{Swallows} M. D. Swallows, T. H. Loftus, W. C. Griffith, B. R. Heckel, E. N. Fortson, and M. V. Romalis, Phys. Rev. A {\bf 87}, 012102 (2013). 
\bibitem{Lamoreaux3} S. K. Lamoreaux, Phys. Rev. A {\bf 53}, (1996).
\bibitem{Lamoreaux4} S. K. Lamoreaux and R. Golub, J. Phys. G: Nucl. Part. Phys. {\bf 36}, 104002 (2009).
\bibitem{Commins} E. D. Commins, Am. J. Phys. {\bf 59}, 1077 (1991).
\bibitem{Lamoreaux2} S. K. Lamoreaux and R. Golub ,Phys. Rev. A {\bf 71}, 032104 (2005).
\bibitem{Pendlebury} J. M. Pendlebury \textit{et al.}, Phys. Rev. A {\bf 70}, 032102 (2004).
\bibitem{Chapman} S. Chapman and T. G. Cowling, \textit{The Mathematical Theory of Non-Uniform Gases}, Cambridge Mathematical Library Series, 1990.
\bibitem{Bello} I. Bello, \textit{Vacuum and Ultravacuum: Physics and Technology}, CRC Press, 2018.
\bibitem{Sandars} P. G. H. Sandars, Phys. Lett. {\bf 22}, 290 (1966).
\bibitem{Flambaum} V. V. Flambaum and I. B. Khriplovich, Zh. Eksp. Teor. Fiz. {\bf 89}, 1505 (1985).
\bibitem{Ginges} J. S. M. Ginges and V. V. Flambaum, Phys. Rep. {\bf 397}, 63 (2004).
\bibitem{Dzuba1} V. A. Dzuba, V. V. Flambaum, and S. G. Porsev, Phys. Rev. A {\bf 80}, 032120 (2009).
\bibitem{Liu} C.-P. Liu, M. J. Ramsey-Musolf, W. C. Haxton, R. G. E. Timmermans, and A. E. L. Dieperink, Phys. Rev. C {\bf 76},  035503 (2007).
\bibitem{Senkov} R. A. Sen'kov, N. Auerbach, V. V. Flambaum, and V. G. Zelevinsky, Phys. Rev. A {\bf 77}, 014101 (2008).
\bibitem{Dzuba2} V. A. Dzuba, V.V. Flambaum, J. S. M. Ginges, and M.G.Kozlov, Phys. Rev. A {\bf 66}, 012111  (2002).
\bibitem{Yoshinaga} N. Yoshinaga, K. Higashiyama, R. Arai, and E. Teruya, Phys. Rev. C {\bf 87}, 044332 (2013).
\bibitem{Dzuba0} V. A. Dzuba, V. V. Flambaum and P. G. Silvestrov, Phys. Lett B {\bf 154}, 93 (1985).
\bibitem{Dmitriev} V. F. Dmitriev, R. A. Sen'kov, and N. Auerbach, Phys. Rev. C {\bf 71}, 035501 (2005).
\bibitem{numrec} W. H. Press \textit{et al.}, {\it Numerical Recipes in C}, (Cambridge University Press, 1992).
\end{thebibliography}
\end{document}